\documentclass[aps,floatfix,amsmath,nofootinbib,amssymb,superscriptaddress]{revtex4}

\usepackage{booktabs} % 引入 booktabs 宏包以使用 \toprule, \midrule, \bottomrule

\usepackage{overpic}
\usepackage{amssymb}
\usepackage{indentfirst}
\usepackage{feynmf}   %{feynmp}
\usepackage{slashed}  %for Feynman symbols
\usepackage{cases}
\usepackage{color}
\usepackage{multirow}
\usepackage{epstopdf}
\usepackage{graphicx,color,bm}
\usepackage{epstopdf}

\usepackage[colorlinks,
            citecolor=green,
            anchorcolor=red,
            menucolor=red,
            linkcolor=red,
            filecolor=red,
            runcolor=red,
            urlcolor=blue,
            frenchlinks=red]{hyperref}

\begin{document}

\title{The $\sin(2\phi)$ azimuthal asymmetry in exclusive $\pi^0$ production}

\author{Chentao Tan}\affiliation{School of Physics, Southeast University, Nanjing
211189, China}

\author{Zhun Lu}
\email{zhunlu@seu.edu.cn}
\affiliation{School of Physics, Southeast University, Nanjing 211189, China}

\begin{abstract}
The $\sin(2\phi)$ azimuthal angular correlation between the transverse momenta of the scattered electron and the recoil proton in the $ep\to e^\prime p^\prime \pi^0$ process provides a probe for quark orbital angular momentum. We numerically calculate this asymmetry for the future Electron-Ion Collider (EIC) in the U.S. and China (EicC) kinematics using a light-front quark-scalar-diquark model, in which the light-front wave functions are derived from the soft-wall AdS/QCD framework. We also investigate the properties of the valence quark angular momentum expressed in terms of helicity-independent and helicity-dependent parton distributions. This study aims to establish theoretical constraints on the asymmetry sensitive to the quark orbital angular momentum prior to its first experimental measurement.

\end{abstract}

%\pacs{12.38.-t, 13.85.Qk, 13.88.+e}
\maketitle

\section{Introduction}

The exploration of the proton's three-dimensional structure is intensifying, with the "proton spin crisis"~\cite{Leader:2013jra} remaining a critical challenge. A key objective is decomposing the proton spin into the spin and orbital angular momentum (OAM) contributions of its constituent quarks and gluons. 
Since the parton spin can be expressed as the first $x$-moment of the corresponding helicity distribution, significant progress has been made in understanding quark and gluon spin contributions (except in the small-$x$ region) through global measurements of parton helicity distributions~\cite{STAR:2014wox,deFlorian:2014yva,Nocera:2014gqa,STAR:2021mqa}, particularly for quarks. Consequently, gluon spin in the small-$x$ region and quark/gluon OAM have become primary targets for future EIC and EicC measurements~\cite{AbdulKhalek:2021gbh,Accardi:2012qut,Anderle:2021wcy}.

In Ref.~\cite{Leader:2013jra}, the authors outline major proton spin decomposition schemes, classified as kinetic (Ji type) and canonical (Jaffe-Manohar type). The kinetic OAM of partons can be expressed as sums of generalized parton distribution (GPD) moments~\cite{Muller:1994ses,Ji:1996ek,Diehl:2003ny,Belitsky:2005qn}, formally equating to the difference between total angular momentum and spin. 
Hard exclusive reactions~\cite{Ji:1996nm,Collins:1998be,Ji:1998pc,Goeke:2001tz,Collins:1996fb} provide access to these GPDs. In contrast, to extract canonical OAM in high-energy scattering, the general OAM definition is introduced as the $\bm{k}_\perp$-moment of the generalized transverse momentum dependent parton distribution (GTMD) $F_{1,4}$~\cite{Lorce:2011kd,Hatta:2011ku,Lorce:2011ni,Mukherjee:2015aja,More:2017zqp}. 
By simply altering the shape of the Wilson line within the correlation functions that define GTMDs~\cite{Meissner:2009ww}, one can derive either the kinetic OAM, linked to a straight Wilson line, or the canonical OAM, associated with a staple-like Wilson line~\cite{Hatta:2011ku,Lorce:2012ce,Ji:2012sj}. 
This distinction arises from assigning the gauge potential term to either the gluon OAM or the quark OAM.
Recent theoretical research has focused on experimental signatures of the gluon GTMD $F_{1,4}^g$ ~\cite{Ji:2016jgn,Hatta:2016aoc,Bhattacharya:2022vvo,Bhattacharya:2018lgm,Boussarie:2018zwg}.
Meanwhile, the quark GTMD $F_{1,4}^q$ within the Efremov-Radyushkin-Brodsky-Lepage (ERBL) region~\cite{Efremov:1978rn,Lepage:1979zb} can be probed via the exclusive double Drell-Yan process~\cite{Bhattacharya:2017bvs}. However, measuring OAM at the forward limit remains a challenge.

Recently, a novel observable directly linked to the quark OAM~\cite{Bhattacharya:2023hbq} was identified for experimental detection of quark GTMD $F^q_{1,4}$ in the Dokshitzer-Gribov-Lipatov-Altarelli-Parisi (DGLAP) region~\cite{Gribov:1972ri,Lipatov:1974qm,Borah:2012ey}.
This process involves exclusive $\pi^0$ production in collisions of unpolarized electrons with longitudinally polarized protons. 
The observable is the average value of the $\sin2(\phi_{l_\perp}-\phi_{\Delta_\perp})$ azimuthal angular correlation between the transverse momentum (${l_\perp}$) of the scattered electron and that of the recoil proton ($\Delta_\perp$), arising from the longitudinal single target-spin asymmetry. 
Given the clean background and the twist-3 contribution unsuppressed by power counting~\cite{Bhattacharya:2023hbq}, this asymmetry is an ideal probe. 
Thus, prospective theoretical calculations using phenomenological models are valuable.

The AdS/CFT correspondence between gravity or string theory in anti-de Sitter (AdS) space and conformal field theories (CFTs) in physical space-time~\cite{Maldacena:1997re,Brodsky:2008pg,Brodsky:2007pt,deTeramond:2011yi} leads to an analytic, semi-classical model for strongly coupled Quantum chromodynamics (QCD), i.e., AdS/QCD, which has scale invariance and dimensional counting at short distances and color confinement at large distances. Although QCD is not conformally invariant, one can use the mathematical representation of the conformal group in five-dimensional AdS space to construct a first approximation to the theory. The correspondence is holographic in the sense that it determines a duality between theories with different numbers of space-time dimensions. In practice, the duality provides an effective gravity description in a $(d+1)$-dimensional AdS space-time in terms of a flat $d$-dimensional conformally invariant quantum field theory defined at the AdS asymptotic boundary~\cite{Gubser:1998bc,Witten:1998qj}. Thus, in principle, one can compute physical observables in a strongly coupled gauge theory in terms of a classical gravity theory. Light-front holography is a remarkable feature of the AdS/CFT correspondence~\cite{Brodsky:2008pf,Brodsky:2008gc,Brodsky:2009xj}, and it allows hadronic amplitudes in the AdS fifth dimension to be precisely mapped to the frame-independent light-front wave functions (LFWFs) of hadrons in physical space-time, thus providing a relativistic description of hadrons at the amplitude level. 

Over the past two decades, the soft-wall AdS/QCD approach~\cite{Karch:2006pv,Brodsky:2006uqa,Andreev:2006vy,Kirsch:2006he}, based on an effective action constructed using a quadratic dilaton field that provides confinement in the infrared (IR) region and breaks both conformal and chiral symmetry, has achieved significant progress in describing and understanding hadron structure (mass spectrum, parton distributions, form factors, thermal properties, etc.)~\cite{Brodsky:2014yha}. The dilaton field $\phi(z)$ has a quadratic dependence on the holographic variable $z$ and is multiplied by the dilaton scale parameter $\kappa$ (of the order of a few hundred MeV): $\phi(z)=\exp(-\kappa^2z^2)$. This scale parameter is a key phenomenological quantity that regulates the strength of the dilaton field, modulates the IR softening of AdS space-time, and fixes the hadron mass spectrum~\cite{Gutsche:2012wb,Gutsche:2019jzh,Gutsche:2011vb,deTeramond:2011yi,Brodsky:2007hb}. One of the main advantages of the soft-wall AdS/QCD is the analytical implementation of quark counting rules~\cite{Matveev:1972gb} in describing hadronic form factors at large $Q^2$ (power scaling)~\cite{Brodsky:2014yha,Matveev:1972gb,Brodsky:2007hb,Abidin:2009hr,Vega:2010ns,Gutsche:2011vb,Gutsche:2019jzh,Branz:2010ub,Brodsky:2011xx,Gutsche:2012bp,Gutsche:2016lrz,Lyubovitskij:2020gjz}. Moreover, this formalism is useful for constraining other approaches~\cite{Gutsche:2013zia,Gutsche:2016gcd}. The quark-diquark model commonly describes protons as composites of an active valence quark and a spectator diquark composed of the remaining two valence quarks, with various parameterizations~\cite{Kroll:1990hg,Jakob:1993th,Jakob:1997wg,Bacchetta:2008af}, which have been used to evaluate quark distributions such as transverse momentum dependent distributions (TMDs), GPDs, and Wigner distributions~\cite{Mondal:2015uha,Vega:2013bxa}, and investigate GPD-TMD relations~\cite{Maji:2015vsa}. 

In this work, we employ a light-front quark-scalar-diquark model~\cite{Gutsche:2016gcd} with LFWFs from soft-wall AdS/QCD, which predicts a general two-particle bound-state wave function~\cite{Brodsky:2007hb}, incorporating valence structure and nonperturbative proton ingredients. Key advantages include LFWFs independent of quark/diquark mass parameters (lacking direct QCD connections), and partial $x$-dependence encoded in unpolarized and helicity-dependent valence quark distributions. 
This model provides an excellent description of experimental data while automatically satisfying fundamental QCD requirements. It precisely reproduces the nucleon magnetic moments and the electromagnetic radii, naturally obeys quark counting rules ($F_1^q\sim 1/Q^4$, $F_2^q\sim 1/Q^6$ at large $Q^2$) and Drell-Yan-West duality, and yields the valence quark parton distribution functions (PDFs) in agreement with global fits, and the Dirac and Pauli form factors and the Sachs nucleon form factors in excellent agreement with experimental data~\cite{Gutsche:2013zia,Gutsche:2016gcd,Gutsche:2014yea}. Furthermore, the derived TMDs, independently on the longitudinal functions $D_q^{(i)}$, satisfy all relations and inequalities found before in theoretical approaches, including the Soffer inequalities and the positivity conditions~\cite{Maji:2015vsa,Lorce:2011zta,Soffer:1994ww,Bacchetta:1999kz,Avakian:2010br}. 
The PDFs are parameterized by fitting at an initial scale $\mu_0=2\,\text{GeV}$. We compute the azimuthal asymmetry, quark form factors, and angular momentum properties.

The paper is organized as follows. In Sec.~\ref{Sec:2},  the kinematics of the exclusive $\pi^0$ production process and the parametrization of the quark GTMDs related to it are introduced. In Sec.~\ref{Sec:3}, the valence quark form factors  and  several angular momentum properties in terms of PDFs are investigated. In Sec.~\ref{Sec:4}, we analytically calculate the transverse moment of GTMDs with nonzero skewness and obtain the numerical results of the azimuthal asymmetry for the EIC and EicC kinematics. In Sec.~\ref{Sec:5}, we summarize this work.

\section{Quark GTMDs in exclusive $\pi^0$ production}\label{Sec:2}

\begin{figure}
	\centering
	\includegraphics[width=0.48\columnwidth]{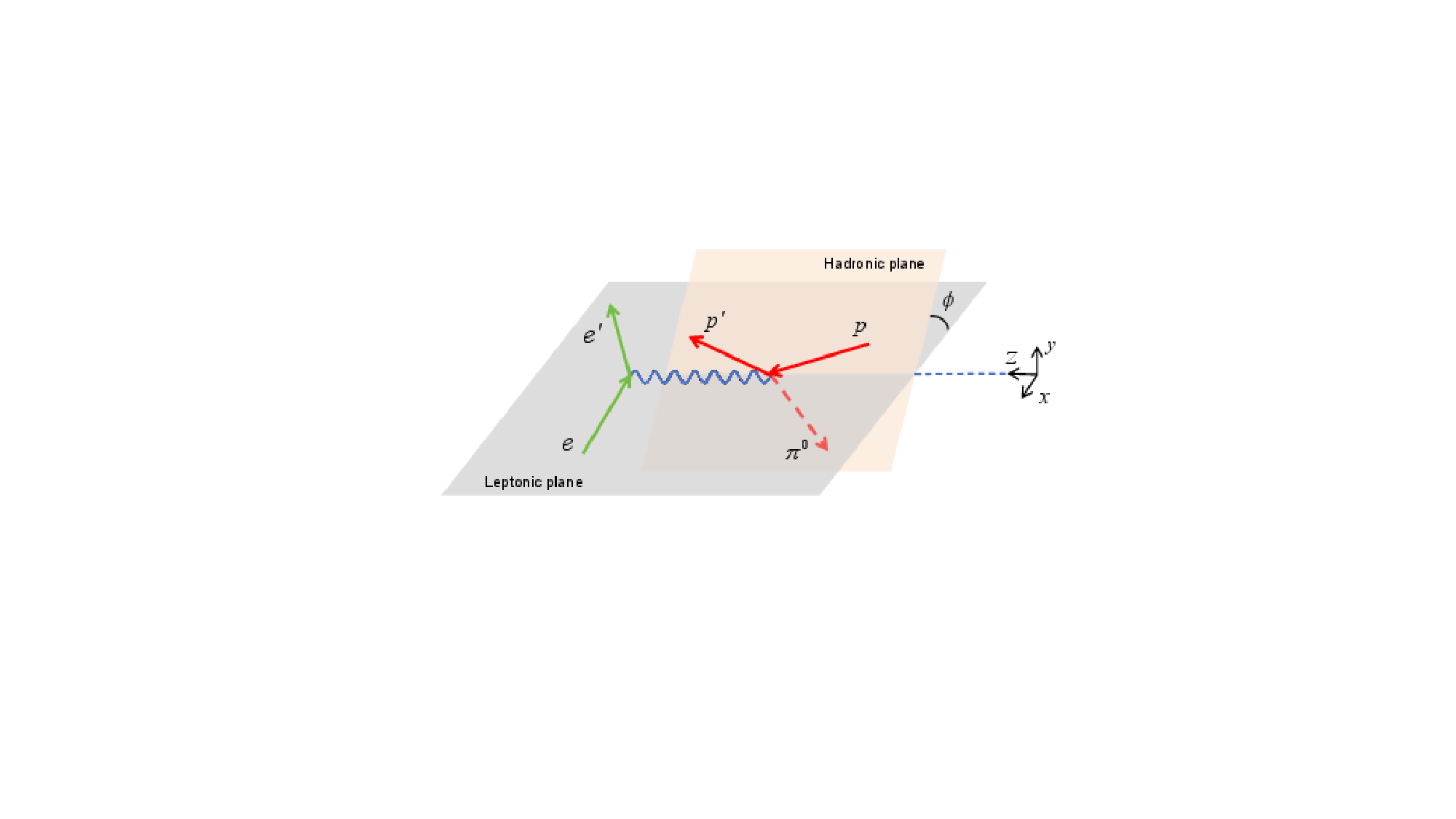}
	\caption{Kinematics of the exclusive $\pi^0$ production in electron-proton collisions.}
	\label{fig:pi0}
\end{figure}

The process we study is the exclusive $\pi^0$ production process in (unpolarized) electron-(longitudinally polarized) proton collisions~\cite{Bhattacharya:2023hbq} 
\begin{align}
	e(l)+p(p,\lambda)\to \pi^0(l_\pi)+e(l^\prime)+p(p^{\prime},\lambda^{\prime}),
\end{align}
as shown in Fig.~\ref{fig:pi0}.
Here $p$ ($p^\prime$) and $\lambda$ ($\lambda^\prime$) denote the momentum and helicity of the initial (final) proton, respectively, $l$ ($l^\prime$) denotes the momentum of the initial (final) electron, and $l_\pi$ the produced $\pi^0$ momentum. 
Following the definition of the light-front vector components, i.e. $v=(v^+,v^-,\bm{v}_\perp)=(\frac{v^0 + v^z}{\sqrt{2}},\frac{v^0 - v^z}{\sqrt{2}}, \bm{v}_\perp)$, they can be expressed as
\begin{align}
	p&=\left(\frac{(1+\xi)P^+}{\sqrt{2}},\frac{M^2+\bm{\Delta}_\perp^2/4}{\sqrt{2}(1+\xi)P^+},
	-\frac{\bm{\Delta}_\perp}{2}\right),
	\label{eq:p}\\ p^{\prime}&=\left(\frac{(1-\xi)P^+}{\sqrt{2}},\frac{M^2+\bm{\Delta}_\perp^2/4}
	{\sqrt{2}(1-\xi)P^+},\frac{\bm{\Delta}_\perp}{2}\right),
	\label{eq:pp}\\
l&=\left(\frac{Q(1-y)}{\sqrt{2}y},\frac{Q}{\sqrt{2}y},\frac{Q\sqrt{1-y}}{y}\right).
\label{eq:l}
\end{align}

Here we neglect the pion mass ($l_\pi^2\approx0$) to simplify calculations. $P=(p^{\prime}+p)/2$ denotes the average momentum of the initial and final protons. 
The skewness variable is defined as $$\xi=(p^{+}-p^{\prime+})/(p^{+}+p^{\prime+})=-\Delta^+/(2P^+)=x_B/(2-x_B),$$ 
with  $x_B=Q^2/2p\cdot q$ the Bjorken variable and $Q^2=-q^2=-(l-l^\prime)^2$. 
The momentum transfer squared is 
$$t=\Delta^2=(p^{\prime}-p)^2=-(4\xi^2M^2+\bm{\Delta}^2_\perp)/(1-\xi^2),$$
where $\bm{\Delta}_\perp$ denotes the
transverse component and $M$ is the proton mass. 
Since we work in the symmetric frame, the transverse momenta of the initial and final protons are given by $\bm{p}_\perp=-\bm{\Delta}_\perp/2$ and $\bm{p}^{\prime}_\perp=\bm{\Delta}_\perp/2$, respectively. 
Finally, $y=p\cdot q/p\cdot l\approx Q^2/x_B\cdot s_{ep}$ represents the lepton momentum loss in the target rest frame.

The validity of the $\sin2(\phi_{l_\perp}-\phi_{\Delta_\perp})$ azimuthal asymmetry as an ideal probe for the quark OAM and its twist-3 origin  have been systematically explained In Ref.~\cite{Bhattacharya:2023hbq}. 
To avoid the dominance of the Primakoff~\cite{Primakoff:1951iae,Liping:2014wbp,Kaskulov:2011ab,Lepage:1980fj,Khodjamirian:1997tk,Jia:2022oyl} and gluon-initiated~\cite{Bhattacharya:2023yvo} processes at extremely small $|t|$, we focus on the valence quark region $\xi\sim 0.1$, where six leading-twist quark GTMDs contribute to the process.  
At fixed light-front time $z^+=0$, the quark GTMDs can be parameterized by the off-forward quark-quark correlator~\cite{Meissner:2009ww}.
\begin{align}	W^{[\Gamma]}_{\lambda^{\prime}\lambda}(x,\xi,\bm{k}_\perp,\bm{\Delta}_\perp)=\frac{1}{2}\int\frac{dz^-d^2\bm{z}_\perp}{(2\pi)^3}e^{ik\cdot z}\langle p^{\prime},\lambda^{\prime}|\bar{\psi}(-z/2)\Gamma\mathcal{W}_{[-z/2,z/2]}\psi(z/2)|p,\lambda\rangle|_{z^+=0},
	\label{eq:WGamma}
\end{align}
where $\mathcal{W}_{[-z/2,z/2]}$ is a Wilson line that ensures the color gauge invariance of the bilocal quark operator. 
$\Gamma$ represents one of the leading twist Dirac $\gamma$-matrices $\{\gamma^+,\gamma^+\gamma_5,i\sigma^{j+}\gamma_5\}$, which corresponds to the unpolarized, longitudinally polarized and transversely polarized quarks, respectively. 

In this work, the parameterizations of Eq.~(\ref{eq:WGamma}) for $\Gamma=\gamma^+,\,\gamma^+\gamma_5$ are relevant . In the notation of~\cite{Meissner:2009ww}, they read 
\begin{align}
	W^{[\gamma^+]}_{{\lambda^{\prime}\lambda}}
	=&\frac{1}{2M}\bar{u}(p^{\prime},\lambda^{\prime})\bigg[F_{1,1}+\frac{i\sigma^{i+}k_\perp^i}{P^+}F_{1,2} +\frac{i\sigma^{i+}\Delta_\perp^i}{P^+}F_{1,3}+\frac{i\sigma^{ij}k_\perp^i\Delta^j_\perp}{M^2}F_{1,4}\bigg]u(p,\lambda) \nonumber\\
	=&\frac{1}{M\sqrt{1-\xi^2}} \bigg\{\bigg[M\delta_{\lambda^{\prime},\lambda}-\frac{1}{2}(\lambda\Delta_\perp^1+i\Delta_\perp^2)\delta_{-\lambda^{\prime},\lambda}\bigg]F_{1,1} +(1-\xi^2)(\lambda k_\perp^1+ik_\perp^2)\delta_{-\lambda^{\prime},\lambda}F_{1,2} \nonumber \\
	&+(1-\xi^2)(\lambda\Delta_\perp^1+i\Delta_\perp^2)\delta_{-\lambda^{\prime},\lambda}F_{1,3} +\frac{i\epsilon_\perp^{ij}k_\perp^i\Delta_\perp^j}{M^2} \bigg[\lambda M \delta_{\lambda^{\prime},\lambda}-\frac{\xi}{2}(\Delta_\perp^1+i\lambda\Delta_\perp^2)\delta_{-\lambda^{\prime},\lambda}\bigg]F_{1,4} \bigg\},
	\label{eq:wgamma+}\\
	W^{[\gamma^+\gamma_5]}_{\lambda^{\prime}\lambda}
	=&\frac{1}{2M}\bar{u}(p^{\prime},\lambda^{\prime})\bigg[-\frac{i\epsilon_\perp^{ij}k_\perp^i\Delta_\perp^j}{M^2} G_{1,1}+\frac{i\sigma^{i+}\gamma_5 k_\perp^i}{P^+}G_{1,2} +\frac{i\sigma^{i+}\gamma_5 \Delta_\perp^i}{P^+}G_{1,3}+i\sigma^{+-}\gamma_5 G_{1,4}\bigg]u(p,\lambda) \nonumber\\
	=&\frac{1}{M\sqrt{1-\xi^2}} \bigg\{-\frac{i\epsilon^{ij}_\perp k_\perp^i\Delta_\perp^j}{M^2} \bigg[M\delta_{\lambda^{\prime},\lambda}-\frac{1}{2}(\lambda\Delta_\perp^1+i\Delta_\perp^2)\delta_{-\lambda^{\prime},\lambda}\bigg]G_{1,1} +(1-\xi^2)(k_\perp^1+i\lambda k_\perp^2)\delta_{-\lambda^{\prime},\lambda}G_{1,2} \nonumber \\
	&+(1-\xi^2)(\Delta_\perp^1+i\lambda\Delta_\perp^2)\delta_{-\lambda^{\prime},\lambda}G_{1,3} +\bigg[\lambda M \delta_{\lambda^{\prime},\lambda}-\frac{\xi}{2}(\Delta^1_\perp+i\lambda\Delta^2_\perp)\delta_{-\lambda^{\prime},\lambda}\bigg]G_{1,4} \bigg\},
	\label{eq:wgamma+5}
\end{align}
GTMDs generally depend on $(x,\xi,\bm{k}_\perp^2,\bm{\Delta}_\perp^2,\bm{k}_\perp\cdot\bm{\Delta}_\perp)$, omitted here for brevity. They have been computed in various models~\cite{Lorce:2011kd,Lorce:2011ni,Mukherjee:2015aja,Lorce:2011dv,Mukherjee:2014nya,
More:2017zqq,Liu:2015eqa,Chakrabarti:2016yuw,Chakrabarti:2017teq,Chakrabarti:2019wjx,Kaur:2019lox,Kumar:2017xcm,Kanazawa:2014nha}, Of particular importance are $F_{1,4}$ and $G_{1,1}$, which characterize nucleon spin structure. 
Besides defining canonical OAM via $F_{1,4}$, both $F_{1,4}$ and $G_{1,1}$ quantify the spin-orbit correlation strength~\cite{Lorce:2011kd,Kanazawa:2014nha,Lorce:2014mxa}: the former encodes the correlation between the nucleon spin and the quark OAM, while the latter the correlation between the quark spin and the quark OAM.
This is because $\mathrm{Re}\, F_{1,4}$ is closely related to the distribution of unpolarized quarks in a longitudinally polarized nucleon with $\mathrm{Re}\, G_{1,1}$ the distribution of longitudinally polarized quarks in an unpolarized nucleon, respectively.

\section{Valence quark PDFs, form factors and properties}\label{Sec:3}

\subsection{Proton LFWFs}

We begin with the light-front quark-scalar-diquark model for protons motivated by the soft-wall AdS/QCD~\cite{Gutsche:2016gcd,Lyubovitskij:2020otz}. 
The two-particle Fock-state expansion for the proton with longitudinal spin $J^z=\pm1/2$ and spin-0 diquark is~\cite{Maji:2016yqo,Maji:2022tog,Maji:2017ill,Maji:2017bcz}
\begin{align} |q\,S\rangle^{\pm}=\int\frac{dxd^2\bm{k}_\perp}{2(2\pi)^3\sqrt{x(1-x)}}
\left[\psi^{\pm}_{+q}(x,\bm{k}_\perp)|
+\frac{1}{2}\lambda_S;xP^+,\bm{k}_\perp\rangle+\psi^{\pm}_{-q}(x,\bm{k}_\perp)|
-\frac{1}{2}\lambda_S;xP^+,\bm{k}_\perp\rangle\right],
\end{align}
where $|\lambda_q\lambda_S;xP^+,\bm{k}_\perp\rangle$ represents the two-particle states consisting of a struck quark with the helicity $\lambda_q=\pm1/2$ and a scalar diquark with the helicity $\lambda_S=0$ (singlet). The LFWFs $\psi_{\lambda_qq}^{\lambda}(x,\bm{k}_\perp)$ defined at the initial scale $\mu_0$ with the proton helicity $\lambda=\pm1/2$ are given by~\cite{Gutsche:2016gcd,Lyubovitskij:2020otz}
\begin{align}
	\psi^+_{+q}(x,\bm{k}_\perp)&=\phi_q^{(1)}(x,\bm{k}_\perp),\nonumber\\
	\psi^+_{-q}(x,\bm{k}_\perp)&=-\frac{k_\perp^1+ik_\perp^2}{M}\phi_q^{(2)}(x,\bm{k}_\perp),\nonumber\\
	\psi^-_{+q}(x,\bm{k}_\perp)&=\frac{k_\perp^1-ik_\perp^2}{M}\phi_q^{(2)}(x,\bm{k}_\perp),\nonumber\\
	\psi^-_{-q}(x,\bm{k}_\perp)&=\phi_q^{(1)}(x,\bm{k}_\perp),
	\label{eq:LFWF}
\end{align}	
where
\begin{align}
	\phi_q^{(1)}(x,\bm{k}_\perp)&=\frac{4\pi}{\kappa}\sqrt{\frac{q_v(x)+\Delta q_v(x)}{2}}\sqrt{D_q^{(1)}(x)}\text{exp}\left[-\frac{\bm{k}_\perp^2}{2\kappa^2}D_q^{(1)}(x)\right],
	\label{eq:phi1}\\
	\frac{1}{M}\phi_q^{(2)}(x,\bm{k}_\perp)&=c_q\frac{4\pi}{\kappa^2}\sqrt{\frac{q_v(x)-\Delta q_v(x)}{2}}D_q^{(2)}(x)\text{exp}\left[-\frac{\bm{k}_\perp^2}{2\kappa^2}D_q^{(2)}(x)\right].
	\label{eq:phi2}
\end{align}
Here, $c_u=1$ and $c_d=-1$; $q_v(x)$ and $\Delta q_v(x)$ are the helicity-independent and helicity-dependent valence quark PDFs, which can be taken from present world data analysis. 
Thus, only the four longitudinal wave functions $D_q^{(i)}(x)$ with $q=u,\,d$ and $i=1,\,2$ need to be determined by the soft-wall AdS/QCD. 
In general, the parameterizations for $D_q^{(1)}(x)$ and $D_q^{(2)}(x)$ are different, since the LFWFs parameterized by them have different combinations of the quark and proton helicities. 
However, they explicitly appear in the flavor contributions to the proton electromagnetic form factors, so the parameters involved can be determined by fitting the  data of the flavor form factors~\cite{Cates:2011pz,Diehl:2013xca}. 
The proton mass $M$ and the dilaton scale parameter $\kappa$ are related in our model by $M=2\sqrt{2}\kappa$~\cite{Gutsche:2011vb,Gutsche:2019jzh,Gutsche:2012wb}, where $\kappa=350\,\text{MeV}$~\cite{Abidin:2009hr,Vega:2010ns}.

The wave functions $\phi_q^{(1)}(x,\bm{k}_\perp) $ and $\phi_q^{(2)}(x,\bm{k}_\perp) $ are generalizations of the results derived by matching the electromagnetic form factors of the proton in the soft-wall AdS/QCD and light-front QCD~\cite{Gutsche:2013zia,Gutsche:2014yea}. 
They are normalized in such a way that their moments combine to produce the unpolarized and helicity PDFs, the number $n_q$, and the axial charge $g_A^q$ of $u$ and $d$ valence quarks in the proton:
\begin{align}	\int\frac{d^2\bm{k}_\perp}{16\pi^3}\left\{\big[\phi_q^{(1)}(x,\bm{k}_\perp)\big]^2
+\frac{\bm{k}_\perp^2}{M^2}\big[\phi_q^{(2)}(x,\bm{k}_\perp)\big]^2\right\}=&q_v(x),\\ \int\frac{d^2\bm{k}_\perp}{16\pi^3}\left\{\big[\phi_q^{(1)}(x,\bm{k}_\perp)\big]^2
-\frac{\bm{k}_\perp^2}{M^2}\big[\phi_q^{(2)}(x,\bm{k}_\perp)\big]^2\right\}=&\Delta q_v(x),\\
	\int^1_0dx\int\frac{d^2\bm{k}_\perp}{16\pi^3}\left\{\big[\phi_q^{(1)}(x,\bm{k}_\perp)\big]^2
+\frac{\bm{k}_\perp^2}{M^2}\big[\phi_q^{(2)}(x,\bm{k}_\perp)\big]^2\right\}=&n_q,
	\label{eq:nq}\\	\int^1_0dx\int\frac{d^2\bm{k}_\perp}{16\pi^3}\left\{\big[\phi_q^{(1)}(x,\bm{k}_\perp)\big]^2
-\frac{\bm{k}_\perp^2}{M^2}\big[\phi_q^{(2)}(x,\bm{k}_\perp)\big]^2\right\}=&g_A^q,
\end{align}
where $n_u=2$ and $n_d=1$.

\subsection{PDFs and form factors}

The valence quark PDFs are defined as the difference between the quark and antiquark components ($q_v(x)=q(x)-\bar{q}(x)$). 
In this work, at the initial scale $\mu_0=2\,\text{GeV}$, we parameterize the unpolarized PDFs using a generic template function~\cite{Martin:2009iq,Cocuzza:2021cbi,Gluck:1998xa,Sufian:2020wcv}:
\begin{align}
	xu_v(x)&=A_ux^{\alpha_u}(1-x)^{\beta_u}(1+\gamma_u\sqrt{x}+\delta_ux),
	\label{eq:xu}\\
	xd_v(x)&=A_dx^{\alpha_d}(1-x)^{\beta_d}(1+\gamma_d\sqrt{x}+\delta_dx),
	\label{eq:xd}
\end{align}
and parameterize the helicity PDFs using a function with as few parameters as possible~\cite{Leader:2001kh}:
\begin{align}
	x\Delta u_v(x)&=B_ux^{a_u}xu_v(x),
	\label{eq:xDu}\\
	x\Delta d_v(x)&=B_dx^{a_d}xd_v(x),
	\label{eq:xDd}
\end{align}
where the normalization constants $A_q$ are given by
the valence number sum rules in Eq.~(\ref{eq:nq}). 
Therefore, there are 6 free parameters ($\alpha_q$, $\beta_q$, $\gamma_q$, $\delta_q$, $B_q$, $a_q$) for the $u$ and $d$ valence quarks, respectively, which are fixed by simultaneously fitting Eq.~(\ref{eq:xu}) and Eq.~(\ref{eq:xDu}) (Eq.~(\ref{eq:xd}) and Eq.~(\ref{eq:xDd})) to the NNPDF4.0 parameterization~\cite{NNPDF:2021njg} for $u_v(x)$ ($d_v(x)$) and the NNPDFpol1.1 parameterization~\cite{Nocera:2014gqa} for $\Delta u_v(x)$ ($\Delta d_v(x)$) at $\mu_0=2\,\text{GeV}$. We select 999 data points in the interval $0.001\leq x\leq 0.999$ with a step size of 0.001. 
The best results with uncertainties for the normalization constants and the parameters from the two fits are listed in Table~\ref{tab1}.

\begin{figure}
	\centering
	\includegraphics[width=0.43\columnwidth]{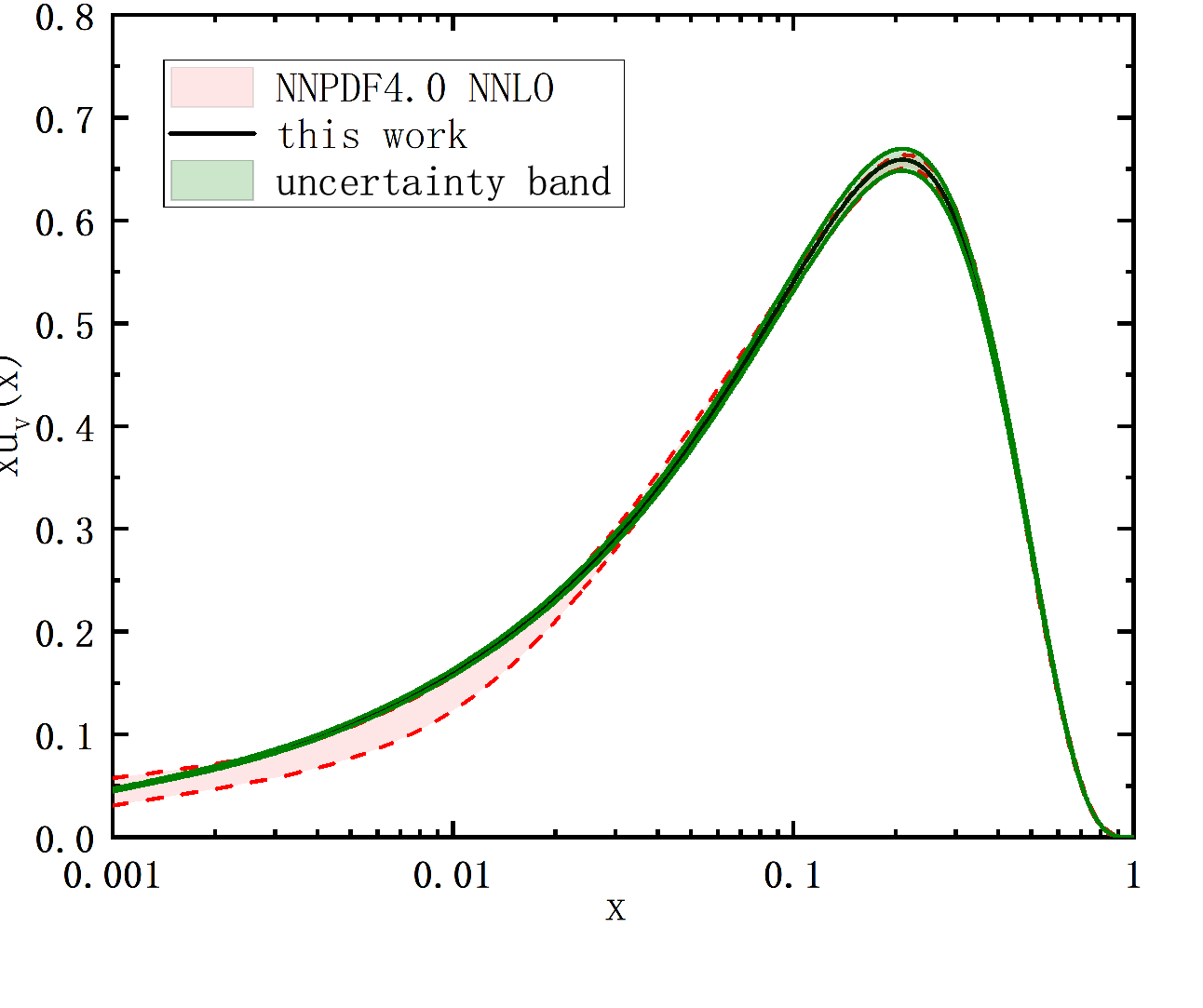}
	\includegraphics[width=0.43\columnwidth]{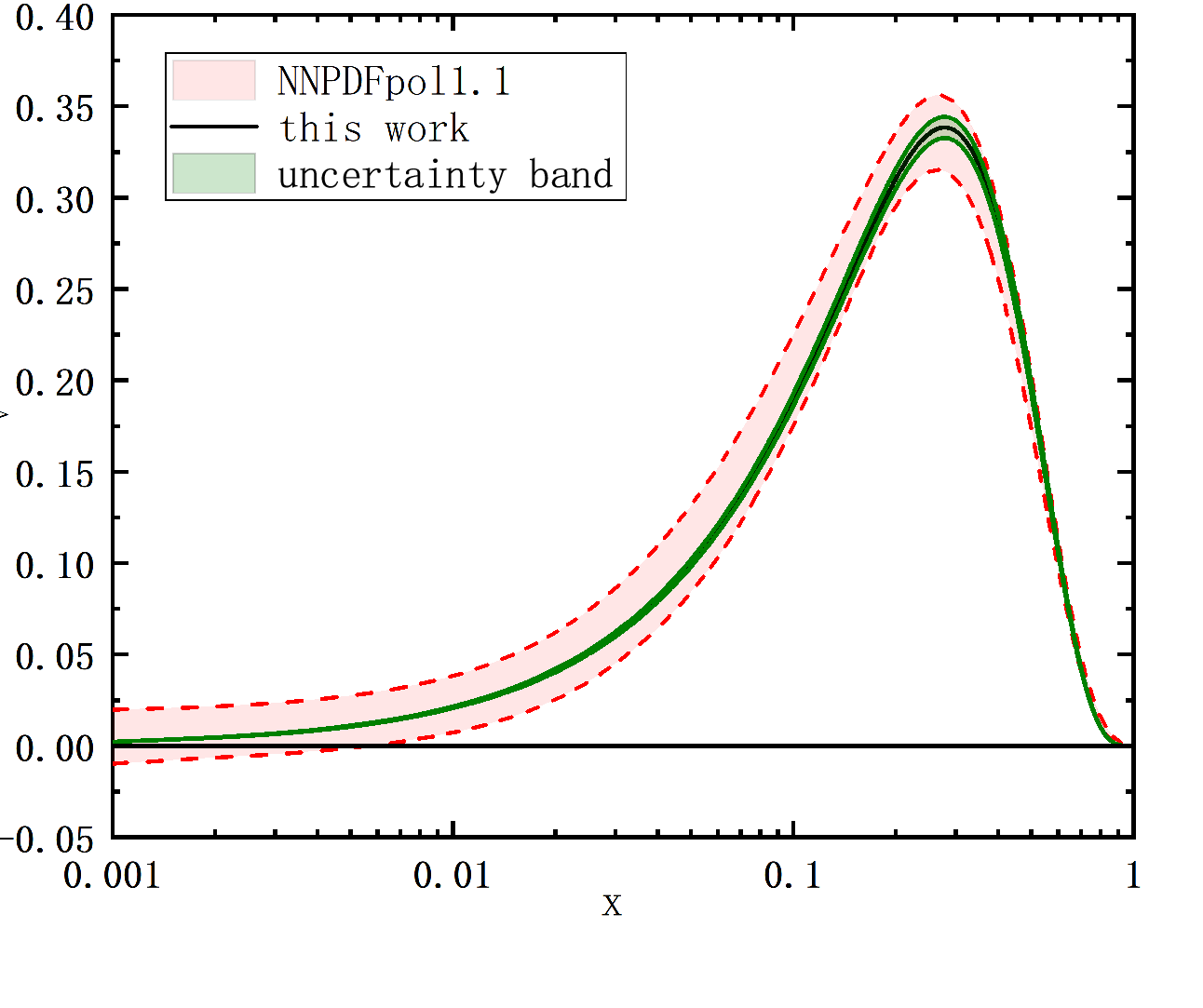}\\
	\includegraphics[width=0.43\columnwidth]{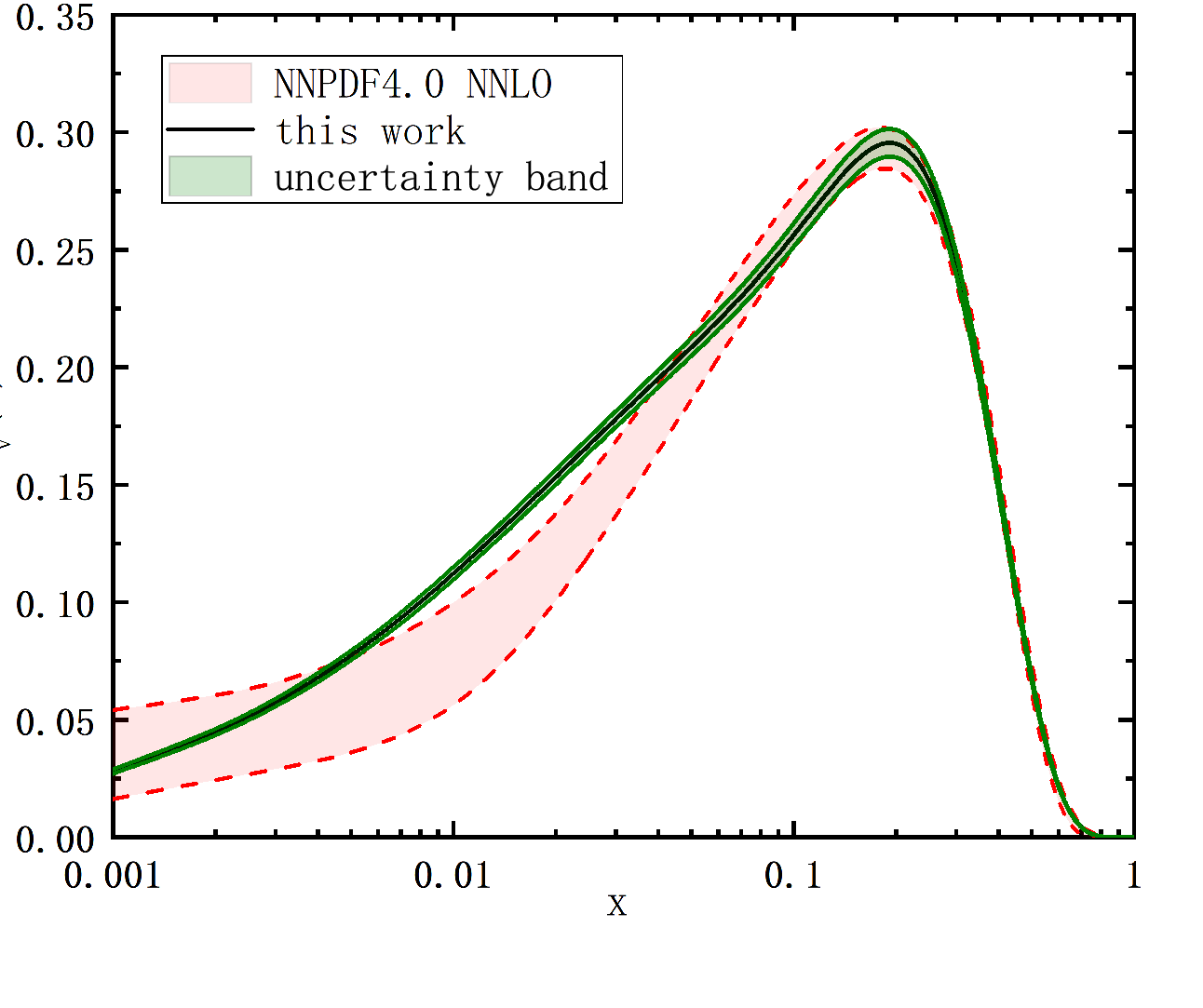}
	\includegraphics[width=0.43\columnwidth]{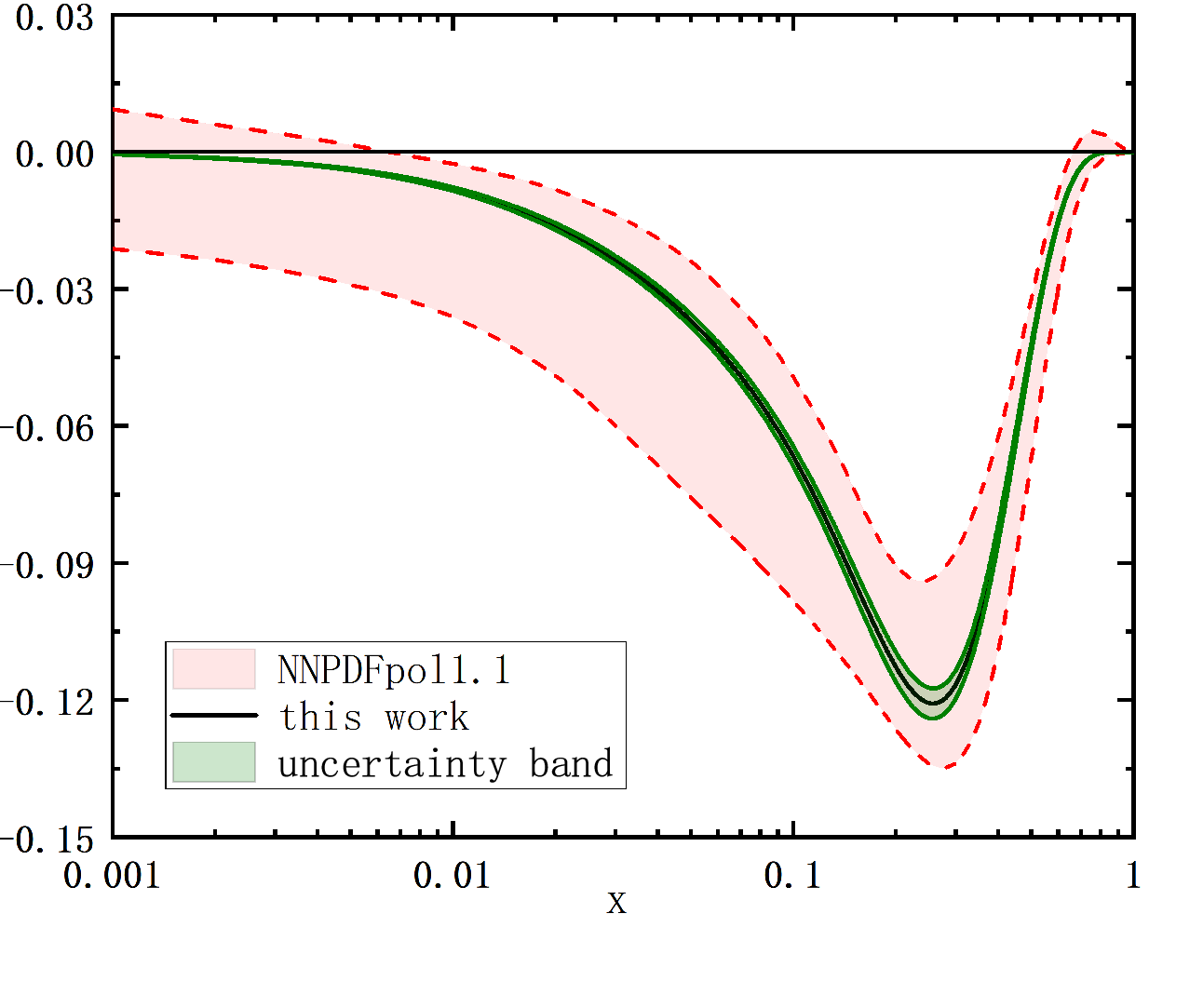}\\
	\caption{The fitted results for $u$ (upper panel) and $d$ (lower panel) valence quark PDFs $xq_v(x)$ (left panel) and $x\Delta q_v(x)$ (right panel), as parameterized in Eqs.~(\ref{eq:xu}-\ref{eq:xDd}).}
	\label{fig:xq}
\end{figure}

\begin{table}[htbp]
	\centering
	\caption{Parameter values of valence quark PDFs}
	\label{tab1}
	\begin{tabular}{lcc} 
		\hline   
		Parameters & $u$ quark ($q=u$) & $d$ quark ($q=d$) \\ 
		\hline 
		~~~~~~$A_q$ & $2.120\pm0.034$  & $5.428\pm0.105$    \\
		~~~~~~$\alpha_q$ & $0.554\pm0.004$ & $0.744\pm0.005$   \\
		~~~~~~$\beta_q$ & $4.159\pm0.006$ & $6.678\pm0.018$   \\
		~~~~~~$\gamma_q$ & $-0.486\pm0.053$ & $-4\pm0.018$   \\
		~~~~~~$\delta_q$ & $5.665\pm0.076$ & $7.938\pm0.058$   \\
		~~~~~~$B_q$ & $0.936\pm0.004$ & $-0.932\pm0.013$   \\
		~~~~~~$a_q$ & $0.427\pm0.003$ & $0.555\pm0.01$   \\
		\hline 
	\end{tabular}
\end{table}

In Fig.~\ref{fig:xq}, we show the results of the simultaneous fits of $xq_v(x)$ (left panel) and $x\Delta q_v(x)$ (right panel) at $\mu_0=2\,\text{GeV}$, where the upper panel corresponds to the $u$ quark and the lower panel corresponds to the $d$ quark. 
The bands with dashed borders represent the NNPDF4.0 parametrization of $xq_v(x)$~\cite{NNPDF:2021njg} and the NNPDFpol1.1 parametrization of $x\Delta q_v(x)$~\cite{Nocera:2014gqa}. 
The bands with solid borders represent the uncertainty bands of our fits. The solid lines represent the results of the center values of parameters. 
In any case, constraining the parameters of $q_v(x)$ and $\Delta q_v(x)$ separately is not enough to reliably describe the $x$-dependence of the model, so it is necessary to perform a simultaneous fit of both the unpolarized and helicity quark PDFs.

The four longitudinal wave functions $D_q^{(i)}(x)$ in Eqs.~(\ref{eq:phi1}-\ref{eq:phi2}), which encode part of the $x$-dependence of LFWFs, can be modeled. 
In Ref.~\cite{Lyubovitskij:2020otz}, $D_q^{(1)}(x)$ is related to the profile function $f_q(x)$ as
\begin{align}
	D_q^{(1)}(x)=-\frac{\text{log}[1-(f_q(x))^{1/2}(1-x)^2]}{(1-x)^2},
	\label{eq:Dq1}
\end{align}
where $f_q(x)$ is determined from the quark PDFs $q_v(x)$ by solving the differential equations with respect to the $x$ variable~\cite{Lyubovitskij:2020otz}:
\begin{align}
	u_v(x)&=[-f_u(x)(1-x)^{4}(1+2\eta_u+(1-x)^2(1-4\eta_u)+2\eta_u(1-x)^4)]^\prime,
	\label{eq:uv}\\ d_v(x)&=\left[-f_d(x)(1-x)^{4}\left(\frac{1}{2}+2\eta_d+(1-x)^2
\left(\frac{1}{2}-4\eta_d\right)+2\eta_d(1-x)^4\right)\right]^\prime,
	\label{eq:dv}
\end{align}
and $D_q^{(2)}(x)$ is defined as~\cite{Gutsche:2016gcd}
\begin{align}
D_q^{(2)}(x)=\sigma_q(x)D_q^{(1)}(x),
\end{align}
where $\sigma_q(x)$ is parameterized as
\begin{align}
	\sigma_q=N_qe^{-\bar{\gamma}_qx}x^{\bar{\alpha}_q} (1-x)^{\bar{\beta}_q}.
	\label{eq:sigma}
\end{align}
The differential equations~(\ref{eq:uv}-\ref{eq:dv}) can be transformed into integral equations, 
\begin{align}
	f_u(x)&=\frac{\int_0^x d\bar{x} \, u_v(\bar{x})-2}{-(1-x)^4[1+2\eta_u+(1-x)^2(1-4\eta_u)+2\eta_u(1-x)^4]},\\
	f_d(x)&=\frac{\int_0^x d\bar{x} \, d_v(\bar{x})-1}{-(1-x)^{4}\left[\frac{1}{2}+2\eta_d+(1-x)^2
		\left(\frac{1}{2}-4\eta_d\right)+2\eta_d(1-x)^4\right]}.
\end{align}
By utilizing the expressions of the quark unpolarized PDFs in Eqs.~(\ref{eq:xu}-\ref{eq:xd}), we obtain the analytic results for the profile functions,
\begin{align}
	f_u(x)=&\frac{-1}{(1-x)^4 \left[ 1 + (1-x)^2 (1-4\eta_u) + 2\eta_u + 2(1-x)^4 \eta_u \right]}\bigg\{\frac{A_u}{1 + \alpha_u + \beta_u}\bigg[ -(1-x)^{1+\beta_u} x^{\alpha_u} \delta_u \nonumber\\
	& + (1 + \alpha_u + \beta_u + \alpha_u \delta_u) \mathrm{B}(x, \alpha_u, 1+\beta_u) + (1 + \alpha_u + \beta_u) \gamma_u \mathrm{B}\bigg(x, \frac{1}{2} + \alpha_u, 1+\beta_u\bigg) \bigg]- 2\bigg\},\\
	f_d(x)=&\frac{-1}{(1-x)^4 \left[ \frac{1}{2} + (1-x)^2 \left(\frac{1}{2} - 4\eta_d\right) + 2\eta_d + 2(1-x)^4 \eta_d \right]}\bigg\{\frac{A_d}{1 + \alpha_d + \beta_d}  \bigg[ -(1-x)^{1+\beta_d} x^{\alpha_d} \delta_d \nonumber\\
	&+ (1 + \alpha_d + \beta_d + \alpha_d \delta_d) \mathrm{B}(x, \alpha_d, 1+\beta_d) + (1 + \alpha_d + \beta_d) \gamma_d \mathrm{B}\bigg(x, \frac{1}{2} + \alpha_d, 1+\beta_d\bigg) \bigg]- 1\bigg\},
\end{align}
where the complete gamma function and the incomplete beta function are defined as
\begin{align}
	\Gamma(z) &= \int_0^\infty dt\, t^{z-1} e^{-t} , \quad \text{Re}(z) > 0,\\
	\mathrm{B}(x;p,q) &= \int_0^x dt \, t^{p-1} (1-t)^{q-1}, \quad 0 \leq x \leq 1.
\end{align}
The results for $D_q^{(1)}(x)$ and $q_v(x)$ in Eqs.~(\ref{eq:Dq1}-\ref{eq:dv}) are obtained at leading twist $\tau=3$. The parameters $\eta_u=2\eta_p+\eta_n$ and $\eta_d=2\eta_n+\eta_p$ are the linear combinations of the nucleon couplings $\eta_N$ with vector field related to the nucleon anomalous magnetic moments $k_N$, where $\eta_N=k_N\kappa/(2\sqrt{2}M)$~\cite{Abidin:2009hr,Vega:2010ns}. Here we adopt the same values as in Ref.~\cite{Lyubovitskij:2020otz}, i.e. $\eta_u=0.211$ and $\eta_d=-1/4$. 

\begin{figure}
	\centering
	\includegraphics[width=0.43\columnwidth]{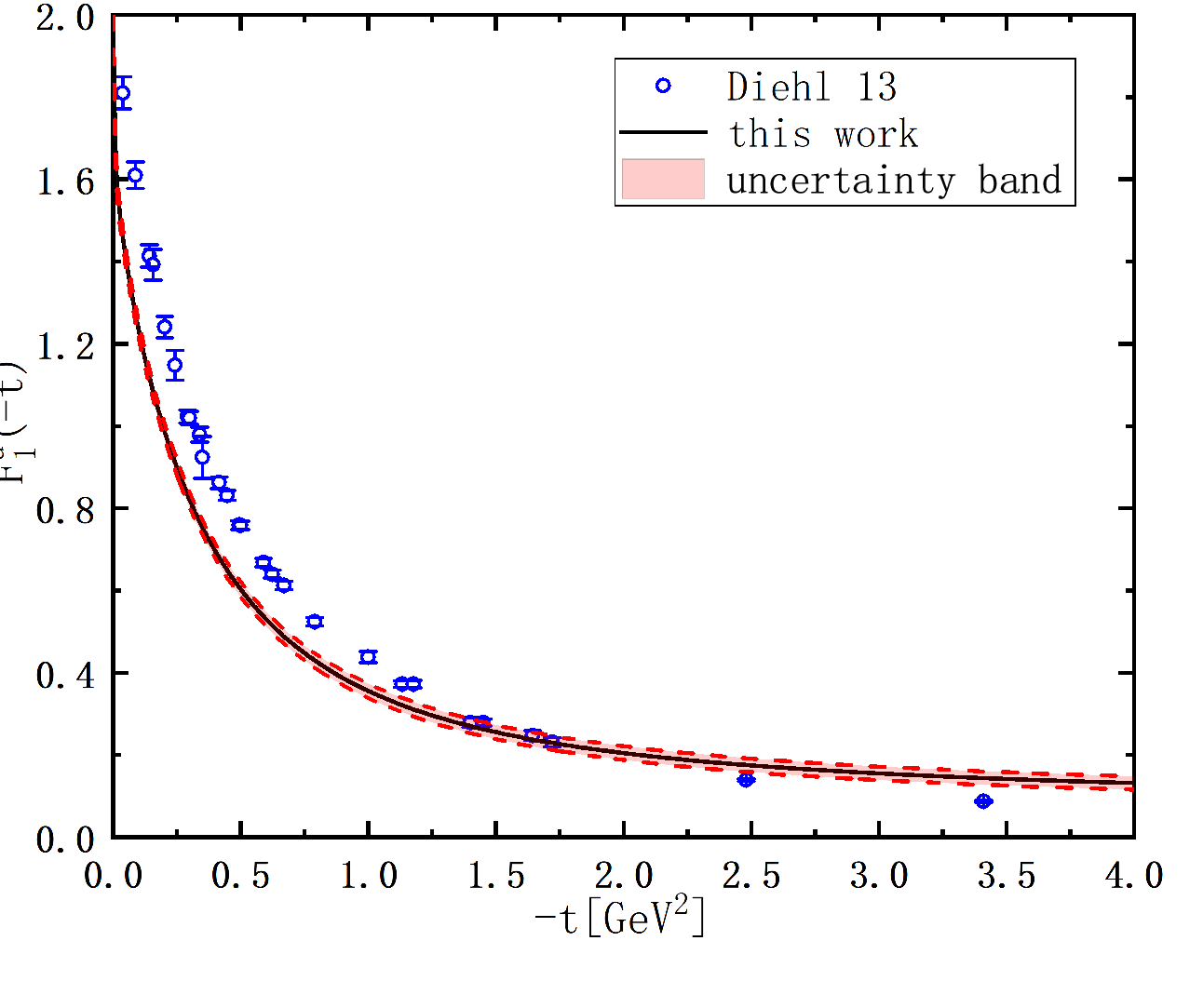}
	\includegraphics[width=0.43\columnwidth]{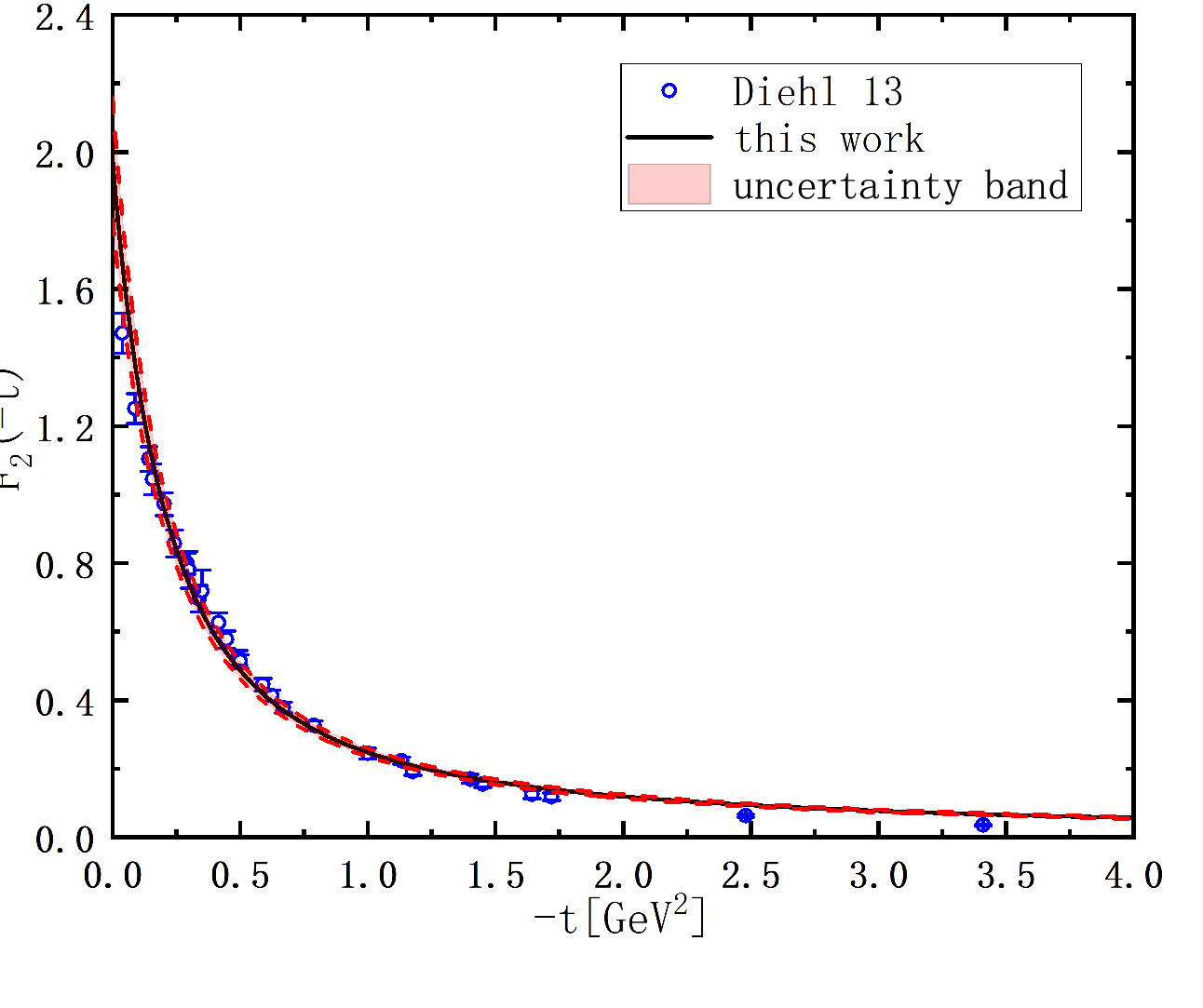}\\
	\includegraphics[width=0.43\columnwidth]{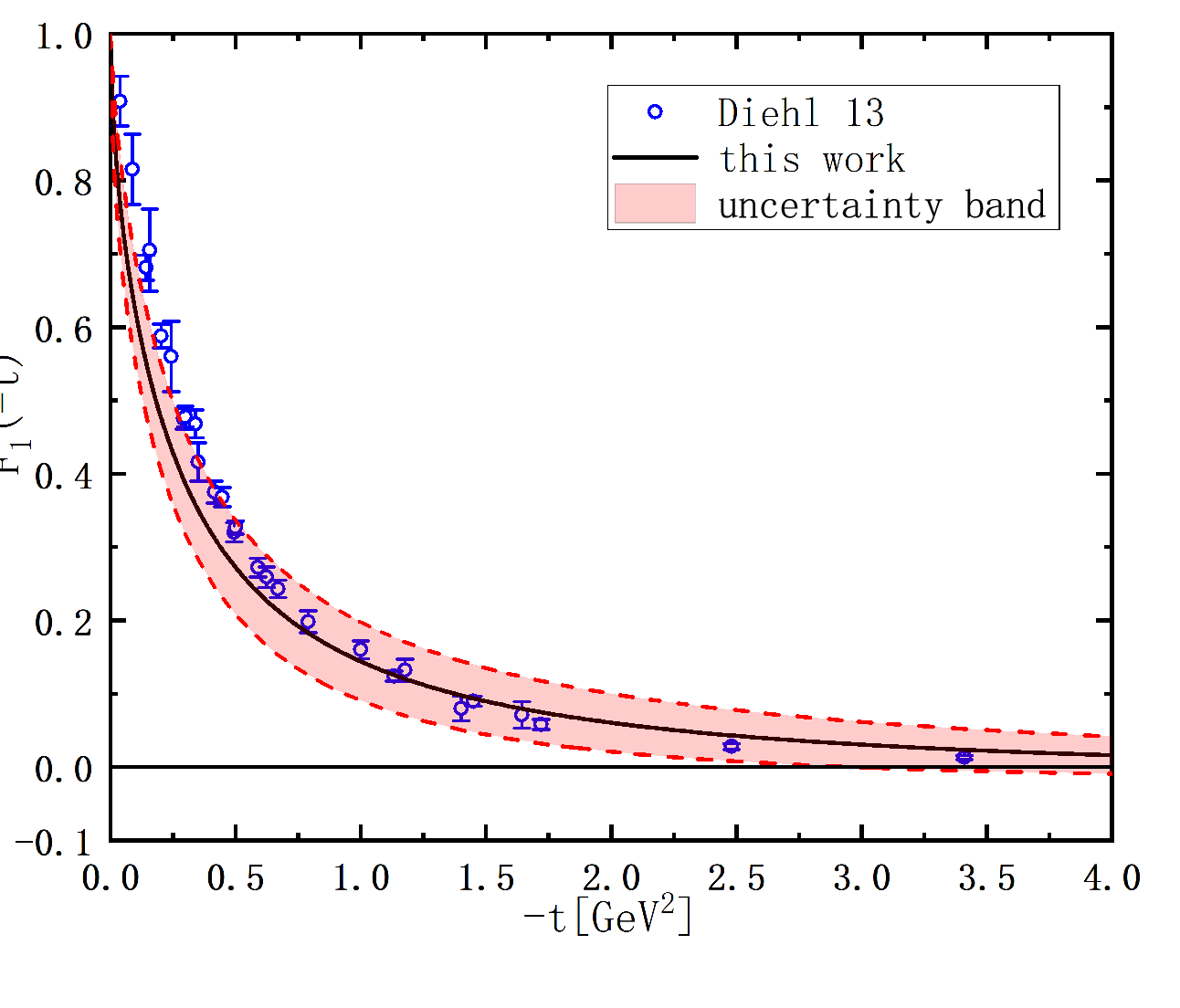}
	\includegraphics[width=0.43\columnwidth]{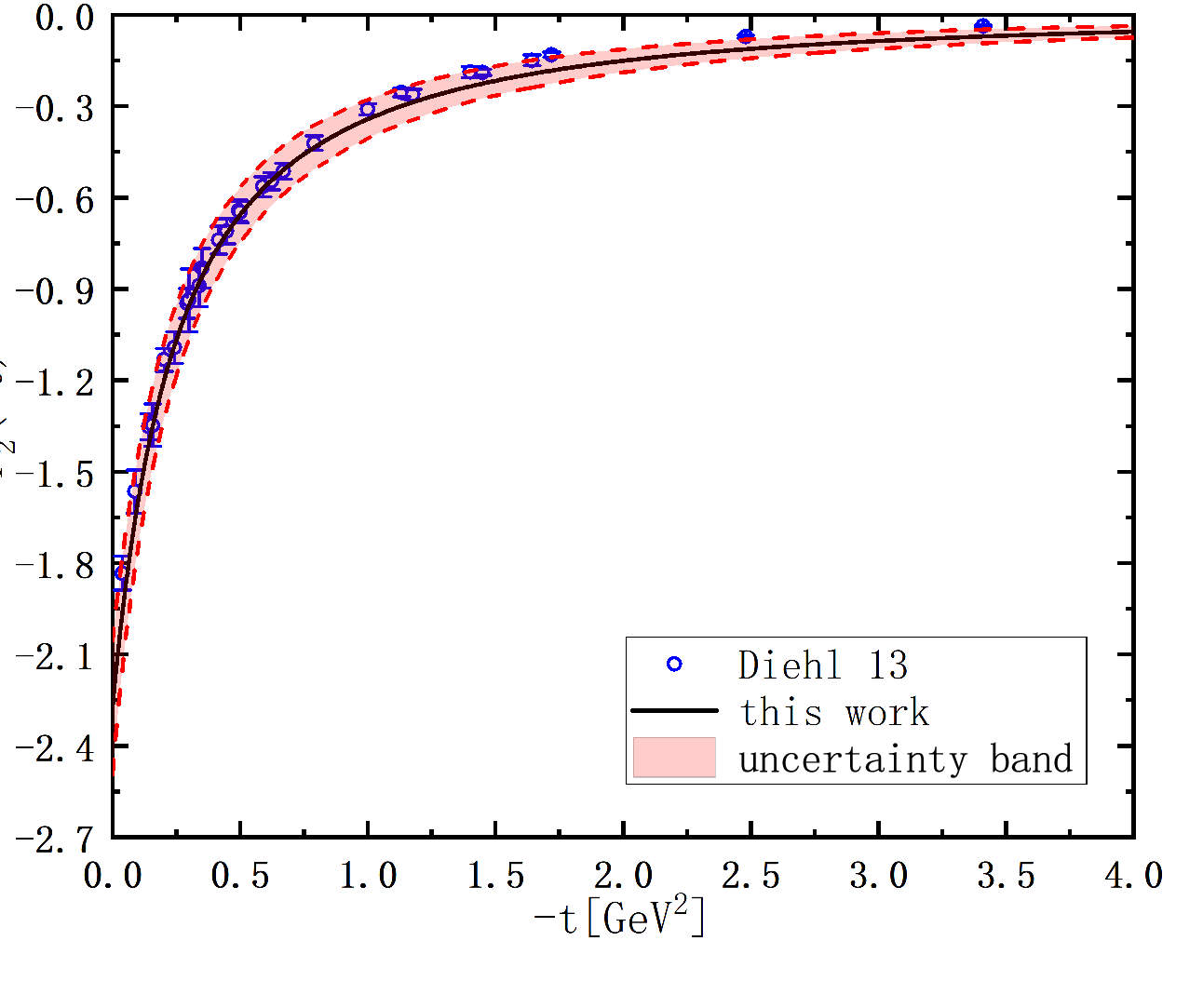}\\
	\caption{The fitted results for $u$ (upper panel) and $d$ (lower panel) valence quark form factors $F_1^q(-t)$ (left panel) and $F_2^q(-t)$ (right panel) in the light-front quark-scalar-diquark model.}
	\label{fig:Fq}
\end{figure}

To obtain the values of the parameters in Eq.~(\ref{eq:sigma}), we express the quark Dirac and Pauli form factors $F^q_{1,2}(-t)$ in terms of LFWFs and fit the analytical results to the data at $\mu_0=2$ GeV from Ref.~\cite{Diehl:2013xca}. 
In the quark-scalar-diquark model, the light-front representation~\cite{Brodsky:2000ii,Brodsky:2000xy} and the corresponding analytic results can be written as
\begin{align} F_1^q(-t)=&\int^1_0dx\int\frac{d^2\bm{k}_\perp}{16\pi^3}\left[\psi^{+\ast}_{+q}(x,\bm{k}^\prime_\perp)
\psi^{+}_{+q}(x,\bm{k}_\perp)+\psi^{+\ast}_{-q}(x,\bm{k}^\prime_\perp)\psi^{+}_{-q}(x,\bm{k}_\perp)\right]\nonumber\\
	=&q_v^+(x)e^{-t_q^{(11)}(x)}+q_v^-(x)e^{-t_q^{(22)}(x)}\big[1-t_q^{(22)}(x)\big],
	\label{eq:F1q}\\ F_2^q(-t)=&-\frac{2M}{\Delta^1_\perp-i\Delta_\perp^2}\int^1_0dx\int\frac{d^2\bm{k}_\perp}{16\pi^3}
\left[\psi^{+\ast}_{+q}(x,\bm{k}^\prime_\perp)
\psi^{-}_{+q}(x,\bm{k}_\perp)+\psi^{+\ast}_{-q}(x,\bm{k}^\prime_\perp)\psi^{-}_{-q}(x,\bm{k}_\perp)\right]\nonumber\\	=&16\sqrt{2}c_q\sqrt{q_v^+(x)q_v^-(x)}\sqrt{D_q^{(1)}(x)}\frac{(1-x)\sigma_q(x)}{[1+\sigma_q(x)]^2}e^{-t_q^{(12)}(x)},
	\label{eq:F2q}
\end{align}
where
\begin{align}
	q_v^{\pm}(x)&=\frac{q_v(x)\pm\Delta q_v(x)}{2},\\ t_q^{(ij)}(x)&=\frac{\bm{\Delta}_\perp^2}{2\kappa^2}\frac{D_q^{(i)}(x)D_q^{(j)}(x)}{D_q^{(i)}(x)+D_q^{(j)}(x)}(1-x)^2,
\end{align}
and $\bm{k}^\prime_\perp=\bm{k}_\perp+\bm{\Delta}_\perp(1-x)$. 
We consider the reference frame where $t=(0,0,\bm{\Delta}_\perp)$ and $-t=\bm{\Delta}_\perp^2>0$. Since constraining the parameters solely through $F_1^q(-t)$ is not enough to reliably describe the $\bm{\Delta}_\perp^2$-behavior of the model, we obtain the best results with uncertainties for the parameters by performing a simultaneous fit of the expressions in Eq.~(\ref{eq:F1q}) and Eq.~(\ref{eq:F2q}) to the data~\cite{Diehl:2013xca} on the flavor decomposition of the nucleon form factors:
\begin{align} N_u&=50.440\pm22.035,\quad\bar{\gamma}_u=38.557\pm3.263,\quad\bar{\alpha}_u=0,\quad
\bar{\beta}_u=-3.891\pm0.329;\nonumber\\
N_d&=0.081\pm0.045,\quad\bar{\gamma}_d=0,\quad\bar{\alpha}_d=-0.820\pm0.174,\quad\bar{\beta}_d=0.485\pm1.153.
\end{align}
Additionally, if any of the four parameters in Eq.~(\ref{eq:sigma}) is set to 0 before performing simultaneous fits, the fitted results will not be as good as those shown in Fig.~\ref{fig:Fq}.

In Fig.~\ref{fig:Fq}, we show the results of the simultaneous fits of $F_1^q(-t)$ (left panel) and $F_2^q(-t)$ (right panel), where the upper panel corresponds to the $u$ quark and the lower panel corresponds to the $d$ quark. 
The circles with error bars represent the data at $\mu_0=2$ GeV taken from Ref.~\cite{Diehl:2013xca}. The bands with red dashed borders represent the uncertainty bands of our fits. The black solid lines represent the results of the center values of parameters. At $t=0$, the quark anomalous magnetic moments: $\kappa_u=F_2^u(0)=1.979\pm0.179$ and $\kappa_d=F_2^d(0)=-2.274\pm0.219$ are $10\%-20\%$ larger than the results in Refs.~\cite{Abidin:2009hr,Vega:2010ns} in absolute terms, where the latter is based on isospin symmetry.

\subsection{Valence quark angular momentum properties}

Based on the above results, we investigate several angular momentum properties of valence quarks. The quark total angular momentum contributing to the proton spin in the Ji decomposition~\cite{Ji:1996ek} is given by
\begin{align}
     J_z^q=&\frac{1}{2}\int^1_0dx x[H_v^q(x,0,0)+E_v^q(x,0,0)]\nonumber\\
     =&\frac{1}{2}\int^1_0dx x[q_v(x)+E_v^q(x)],
\end{align}
and the quark internal spin $S_z^q$ is given by
\begin{align}
	S_z^q=\frac{1}{2}\int^1_0dx\tilde{H}_v^q(x,0,0)=\frac{1}{2}\int^1_0dx \Delta q_v(x).
\end{align}
Thus, we obtain the quark kinetic OAM
\begin{align}
	L_z^q=&J_z^q-S_z^q\nonumber\\
	=&\frac{1}{2}\int^1_0dx\{x[H_v^q(x,0,0)+E_v^q(x,0,0)]-\tilde{H}_v^q(x,0,0)\}\nonumber\\
	=&\frac{1}{2}\int^1_0dx\{x[q_v(x)+E_v^q(x)]-\Delta q_v(x)\}.
\end{align}

In addition, the quark canonical OAM is defined by the Wigner distribution $\rho_{LU}(\bm{b}_\perp,\bm{k}_\perp,x)$ (or $\bm{k}_\perp$-moment of the GTMD $F_{1,4}(x,0,\bm{k}^2_\perp,0,0)$)~\cite{Lorce:2011ni}, which represents the distribution of an unpolarized quark in a longitudinally polarized proton:
\begin{align}
	l_z^q=&\int^1_0dx\int d^2\bm{k}_\perp d^2\bm{b}_\perp(\bm{b}_\perp\times\bm{k}_\perp)_z\rho_{LU}(\bm{b}_\perp,\bm{k}_\perp,x)\nonumber\\
	=&-\int^1_0dx\int d^2\bm{k}_\perp\frac{\bm{k}_\perp^2}{M^2}F_{1,4}(x,0,\bm{k}^2_\perp,0,0)\nonumber\\
	=&\frac{1}{2}\int^1_0dx(1-x)(q_v(x)-\Delta q_v(x)).
\end{align}

Another relevant property is the quark spin-orbit correlation, which is defined by the Wigner distribution $\rho_{UL}(\bm{b}_\perp,\bm{k}_\perp,x)$ (or the $\bm{k}_\perp$-moment of the GTMD $G_{1,1}(x,0,\bm{k}^2_\perp,0,0)$) representing the distribution of a longitudinally polarized quark in an unpolarized proton~\cite{Lorce:2011kd}:
\begin{align}
		C_z^q=&\int^1_0dx\int d^2\bm{k}_\perp d^2\bm{b}_\perp(\bm{b}_\perp\times\bm{k}_\perp)_z\rho_{UL}(\bm{b}_\perp,\bm{k}_\perp,x)\nonumber\\
		=&\int^1_0dx\int d^2\bm{k}_\perp\frac{\bm{k}_\perp^2}{M^2}G_{1,1}(x,0,\bm{k}^2_\perp,0,0)\nonumber\\
		=&-\frac{1}{2}\int^1_0dx(1-x)(q_v(x)-\Delta q_v(x)).
\end{align}
We find that it is opposite to the canonical OAM $l_z^q$:
\begin{align}
	C_z^q=-l_z^q,
\end{align}
due to $\rho_{UL}(\bm{b}_\perp,\bm{k}_\perp,x)=-\rho_{LU}(\bm{b}_\perp,\bm{k}_\perp,x)$ (or $G_{1,1}(x,0,\bm{k}^2_\perp,0,0)=F_{1,4}(x,0,\bm{k}^2_\perp,0,0)$), a general feature of the quark-scalar-diquark model~\cite{Kanazawa:2014nha}. The relation between GTMDs will be confirmed in Sec.~{\ref{Sec:4}}.

Finally, the valence quark number $n_q$ and axial charge $g_A^q$ can also be expressed in terms of the Wigner distributions (or $\bm{k}_\perp$-integrated GTMDs) as follows~\cite{Lorce:2011kd}
\begin{align}
	n_q=&\int^1_0dx\int d^2\bm{b}_\perp d^2\bm{k}_\perp \rho_{UU}(x,\bm{b}_\perp,\bm{k}_\perp)\nonumber\\
	=&\int^1_0dx\int d^2\bm{k}_\perp F_{1,1}(x,0,\bm{k}_\perp^2,0,0),\\
	g_A^q=&\int^1_0dx\int d^2\bm{b}_\perp d^2\bm{k}_\perp \rho_{LL}(x,\bm{b}_\perp,\bm{k}_\perp)\nonumber\\
	=&\int_0^1dx\int d^2\bm{k}_\perp G_{1,4}(x,0,\bm{k}_\perp^2,0,0),
\end{align}
where $\rho_{UU}(\bm{b}_\perp,\bm{k}_\perp,x)$ represents the distribution of an unpolarized quark in an unpolarized proton, and $\rho_{LL}(\bm{b}_\perp,\bm{k}_\perp,x)$ represents the distribution of a longitudinally polarized quark in a longitudinally polarized proton.

\begin{table}[htbp]
	\centering
	\caption{Valence quark angular momentum properties}
	\label{tab2}
	\begin{tabular}{lccc} 
		\hline   
		Properties & LCCQM~\cite{Lorce:2011kd} & LC$\mathcal{X}$QSM~\cite{Lorce:2011kd} & This work \\ 
		\hline 
		~~~~~$J_z^u$ & 0.569 & 0.566 & $0.232\pm0.008$  \\
		~~~~~$J_z^d$ & $-0.069$ & $-0.066$ & $-0.029\pm0.018$  \\
		~~~~~$S_z^u$ & 0.498 & 0.574 & $0.348\pm0.003$  \\
		~~~~~$S_z^d$ & $-0.124$ & $-0.143$ & $-0.117\pm0.003$  \\
		~~~~~$L_z^u$ & 0.071 & $-0.008$ & $-0.116\pm0.009$  \\
		~~~~~$L_z^d$ & 0.055 & 0.077 & $0.088\pm0.018$  \\
		~~~~~$l_z^u$ & 0.131 & 0.073 & $0.584\pm0.003$  \\
		~~~~~$l_z^d$ & $-0.005$ & $-0.004$ & $0.539\pm0.002$  \\
		~~~~~$C_z^u$ & 0.227 & 0.130 & $-0.584\pm0.003$  \\
		~~~~~$C_z^d$ & 0.187 & 0.109 & $-0.539\pm0.002$  \\
		\hline 
	\end{tabular}
\end{table}
In Table~\ref{tab2}, we list the numerical results of the valence quark angular momentum properties ($J_z^q$, $L_z^q$, $S_z^q$, $l_z^q$, $C_z^q$) and compare them with those from the light-cone constituent quark model (LCCQM) and the light-cone version of the chiral quark-soliton model (LC$\mathcal{X}$QSM)~\cite{Lorce:2011kd}. 
We find that most of the results show significant differences, which arise from the differences in the magnetization PDFs $E_v^q(x)$ related to the quark anomalous magnetic moments $\kappa_q$ and the helicity PDFs $\Delta q_v(x)$ related to the quark internal spin $S_z^q$ among different models. 
It is worth noting that in our model, $\Delta q_v(x)$ are obtained from fitting to the NNPDFpol1.1 parametrization, and the parameters of $E_v^q(x)$ are determined through the fitting of flavor form factors.

\section{The $\sin(2\phi)$ azimuthal asymmetry}\label{Sec:4}

In Ref.~\cite{Bhattacharya:2023hbq}, the cross section for the exclusive $\pi^0$ production process in (unpolarized) electron-(longitudinally polarized) proton collisions has been derived as
\begin{align} \frac{d\sigma_T}{dtdQ^2dx_Bd\phi}=&\frac{(N_c^2-1)^2\alpha^2_{em}\alpha_s^2f_\pi^2\xi^3\bm{\Delta}^2_\perp}
{2N_c^4(1-\xi^2)Q^{10}(1+\xi)}[1+(1-y)^2]\nonumber\\ &\times\bigg\{\left[|\mathcal{F}_{1,1}+\mathcal{G}_{1,1}|^2+|\mathcal{F}_{1,4}+\mathcal{G}_{1,4}|^2
+2\frac{M^2}{\bm{\Delta}_\perp^2}|\mathcal{F}_{1,2}+\mathcal{G}_{1,2}|^2\right]\nonumber\\
	&+\text{cos}(2\phi)a[-|\mathcal{F}_{1,1}+\mathcal{G}_{1,1}|^2+|\mathcal{F}_{1,4}+\mathcal{G}_{1,4}|^2]\nonumber\\
	&+\lambda \text{sin}(2\phi)2a \text{Re}[(i\mathcal{F}_{1,4}+i\mathcal{G}_{1,4})(\mathcal{F}^\ast_{1,1}+\mathcal{G}^\ast_{1,1})]\bigg\},
\end{align}
where $N_c=3$ is the number of quark colors, $\alpha_{em}=1/137$ is the running coupling constant of electromagnetic interaction, $\alpha_s$ is the running coupling constant of strong interaction and is set to 0.3, $f_\pi=131\,\text{MeV}$ is the $\pi^0$ decay constant, $\phi=\phi_{l_\perp}-\phi_{\Delta_\perp}$, and $a=2(1-y)/[1+(1-y)^2]$. The notations $\mathcal{F}_{1,m}$ and $\mathcal{G}_{1,m}$ with $m=1,\,2,\,4$ represent the convolutions involving the GTMDs $F_{1,m}$, $G_{1,m}$ and the $\pi^0$ distribution amplitude (DA) $\phi_\pi(z)$, which are specified as
\begin{align}	\mathcal{F}_{1,1}&=\int^1_{-1}dx\int^1_0dz\tilde{F}_{1,1}^{(0)}(x,\xi,\bm{\Delta}_\perp)x^2
\frac{1+z^2-z}{z^2(1-z)^2}\phi_\pi(z),
	\label{eq:mF11}\\ \mathcal{F}_{1,2}&=\int^1_{-1}dx\int^1_0dz\tilde{F}_{1,2}^{(1)}(x,\xi,\bm{\Delta}_\perp)x\xi(1-\xi^2)
\frac{1+z^2-z}{z^2(1-z)^2}\phi_\pi(z),\\ \mathcal{F}_{1,4}&=\int^1_{-1}dx\int^1_0dz\tilde{F}_{1,4}^{(1)}(x,\xi,\bm{\Delta}_\perp)x\xi
\frac{1+z^2-z}{z^2(1-z)^2}\phi_\pi(z),\\	\mathcal{G}_{1,1}&=\int^1_{-1}dx\int^1_0dz\tilde{G}_{1,1}^{(1)}(x,\xi,\bm{\Delta}_\perp)
\frac{x^2+2x^2z+\xi^2}{z^2}\phi_\pi(z),\\ \mathcal{G}_{1,2}&=\int^1_{-1}dx\int^1_0dz\tilde{G}_{1,2}^{(1)}(x,\xi,\bm{\Delta}_\perp)(1-\xi^2)
\frac{x^2+2x^2z+\xi^2}{z^2}\phi_\pi(z),\\
	\mathcal{G}_{1,4}&=\int^1_{-1}dx\int^1_0dz\tilde{G}_{1,4}^{(0)}(x,\xi,\bm{\Delta}_\perp)
\frac{x}{\xi}\frac{4\xi^2z+\xi^2-2x^2z+x^2}{z^2}\phi_\pi(z),
	\label{eq:mG14}
\end{align}
where
\begin{align}
	\tilde{f}_{1,m}^{(n)}(x,\xi,\bm{\Delta}_\perp)=\int d^2\bm{k}_\perp\left(\frac{\bm{k}_\perp^2}{M^2}\right)^n\frac{\frac{1}{\sqrt{2}}
\left(\frac{2}{3}f_{1,m}^u+\frac{1}{3}f_{1,m}^d\right)}{(x+\xi-i\epsilon)^2(x-\xi+i\epsilon)^2},
	\label{eq:f} 
\end{align}
with $n=0,\,1$. $f_{1,m}^u$ and $f_{1,m}^d$ denote the GTMDs of $u$ and $d$ quarks, respectively. To ensure direct comparability between our numerical results and the theoretical predictions in Ref.~\cite{Bhattacharya:2023hbq}, facilitate verification of the feasibility of this observable channel, and validate the physical interpretation of the asymmetry signal, the pion DA $\phi_\pi(z)$ is taken as the asymptotic form $6z(1-z)$, where $z$ denotes the longitudinal momentum fraction of $\pi^0$ carried by the outgoing quark. This choice allows our calculations to clearly demonstrate the contributions of the soft-wall AdS/QCD proton wave functions to the asymmetry without being confounded by model differences in the pion DA. Furthermore, most of the existing experimental measurements and the theoretical calculations of the exclusive $\pi^0$ production use the asymptotic pion DA as the standard input. Adopting this widely accepted form enables our predictions to be directly compared with existing experimental data and other theoretical results, which is crucial for future experimental tests of this signal at the EIC and EicC.

For the quadruple integrals in Eq.~(\ref{eq:mF11}-\ref{eq:mG14}), both the integrals over $x$ and $z$ may diverge. 
On the one hand, the singularities in the $z$-integrated hard parts occur at the endpoints, i.e., $z$ approaches 0 and 1. 
They can be regularized by introducing the transverse momentum dependence of the pion DA~\cite{Goloskokov:2005sd,Goloskokov:2006hr,Sun:2021gmi} to adjust the integration limits from $\int^1_0dz$ to $\int_{\langle \bm{p}_\perp^2\rangle/Q^2}^{1-{\langle \bm{p}_\perp^2\rangle/Q^2}}dz$~\cite{Goloskokov:2007nt}, where $\langle \bm{p}_\perp^2\rangle$ represents the mean-squared transverse momentum of the quark inside the pion and is fixed as $0.04\,\text{GeV}^2$~\cite{Bhattacharya:2023hbq}. 
On the other hand, the interplay between the singularities of quark GTMDs at $x=\pm\xi$ and the double poles at $x=\pm\xi$ in Eq.~(\ref{eq:f}) also leads to divergence, which can be addressed by shifting the double poles from $1/(x\mp\xi\pm i\epsilon)^2$ to $1/(x\mp\xi\mp\langle \bm{p}_\perp^2\rangle/Q^2\pm i\epsilon)^2$~\cite{Anikin:2002wg}. The integration range $[-1,1]$ for $x$ can be divided into three regions~\cite{Lorce:2011dv,Kaur:2019jow,Kaur:2019kpi}: 
(i) the DGLAP region $-1\leq x<-\xi$ for the distributions of antiquarks, 
(ii) the ERBL region $-\xi\leq x\leq \xi$ for the distributions of quark-antiquark pairs, 
and (iii) the DGLAP region $\xi<x\leq1$ for the distributions of quarks. 
Since this process is sensitive to the GTMD $F_{1,4}$ in the DGLAP region and our model is restricted to the three-(valence) quark Fock sector, we only need to consider the region (iii) corresponding to the quark distributions. Then we can perform the integrations over $x$ using the relation~\cite{Nekrasov:2003er,Maina:1993uq}
\begin{align}
	\int^1_\xi dx\frac{f(x)}{(x-\xi-\langle \bm{p}_\perp^2\rangle/Q^2+i\epsilon)^2}=&P.V.\int_\xi^1dx\frac{f(x)}{(x-\xi-\langle \bm{p}_\perp^2\rangle/Q^2)^2}-i\pi\int_\xi^1dxf(x)\delta^\prime(x-\xi-\langle \bm{p}_\perp^2\rangle/Q^2)\nonumber\\
	=&\lim_{\epsilon\to 0}\left[\int_\xi^{\xi+\frac{\langle \bm{p}_\perp^2\rangle}{Q^2}-\epsilon}dx\frac{f(x)-f(\xi+\langle \bm{p}_\perp^2\rangle/Q^2)}{(x-\xi-\langle \bm{p}_\perp^2\rangle/Q^2)^2}+\int_{\xi+\frac{\langle \bm{p}_\perp^2\rangle}{Q^2}+\epsilon}^1dx\frac{f(x)-f(\xi+\langle \bm{p}_\perp^2\rangle/Q^2)}{(x-\xi-\langle \bm{p}_\perp^2\rangle/Q^2)^2}\right]\nonumber\\
	&+f(\xi+\langle \bm{p}_\perp^2\rangle/Q^2)\frac{\xi-1}{(1-\xi-\langle \bm{p}_\perp^2\rangle/Q^2)\langle \bm{p}_\perp^2\rangle/Q^2}+i\pi f^\prime(x)|_{x=\xi+\langle \bm{p}_\perp^2\rangle/Q^2},
\end{align}
where $P.V.$ represents the (Cauchy) principal value prescription. The numerical results of the asymmetry in Ref.~\cite{Bhattacharya:2023hbq} were derived by neglecting contributions from the GTMD $G_{1,1}$ and the real parts of the $\mathcal{F}_{1,2}$, $\mathcal{F}_{1,4}$ and $\mathcal{G}_{1,4}$ terms. 
In this work, we will include both the real and imaginary parts of all convolution terms in numerical calculations.

\subsection{The transverse moments of GTMDs with nonzero skewness in the light-front model}
In this light-front quark-scalar-diquark model, the overlap representations for the correlator in Eq.~(\ref{eq:WGamma}) can be expressed in terms of the proton LFWFs constructed from the soft-wall AdS/QCD prediction in Eq.~(\ref{eq:LFWF}) as~\cite{Maji:2022tog}
\begin{align}	W^{[\gamma^+]}_{\lambda^{\prime}\lambda}&=\frac{1}{16\pi^3}\sum_{\lambda_q}
\psi_{\lambda_qq}^{\lambda^{\prime}\dagger}(x^{\prime\prime},\bm{k}^{\prime\prime}_\perp)
\psi_{\lambda_qq}^{\lambda}(x^{\prime},\bm{k}_\perp^\prime),\\ W^{[\gamma^+\gamma_5]}_{\lambda^{\prime}\lambda}&=\frac{1}{16\pi^3}\sum_{\lambda_q}(2\lambda_q)
\psi_{\lambda_qq}^{\lambda^{\prime}\dagger}(x^{\prime\prime},\bm{k}_\perp^{\prime\prime})
\psi_{\lambda_qq}^{\lambda}(x^{\prime},\bm{k}_\perp^{\prime}).
\end{align}
The initial and final momenta ($\bm{k}_\perp^\prime$ and $\bm{k}_\perp^{\prime\prime}$) and longitudinal momentum fractions ($x^\prime$ and $x^{\prime\prime}$) carried by the struck quark in the DGLAP region $\xi<x\leq 1$ are given by
\begin{align}
	\bm{k}^{\prime}_\perp=&\bm{k}_\perp - (1-x^{\prime})\frac{\bm{\Delta}_\perp}{2},\qquad \text{with} \qquad x^{\prime}=\frac{x+\xi}{1+\xi}, \\
   \text{and}\quad \bm{k}^{\prime\prime}_\perp=&\bm{k}_\perp +(1-x^{\prime\prime}) \frac{\bm{\Delta}_\perp}{2},\qquad \text{with} \qquad x^{\prime\prime}=\frac{x-\xi}{1-\xi},
\end{align}
respectively.

Using the parameterizations in Eqs.~(\ref{eq:wgamma+}-\ref{eq:wgamma+5}) and the proper helicity combinations of quarks and protons, we can express the quark GTMDs in terms of the correlators $W_{\lambda^\prime\lambda}^{[\Gamma]}$. Here we further calculate the transverse moments of these GTMDs in Eqs.~(\ref{eq:mF11}-\ref{eq:mG14}), and the corresponding explicit expressions are:
\begin{align} F_{1,1}^{q(0)}=&2\sqrt{1-\xi^2}\bigg\{\frac{\sqrt{D_q^{(1)}(x^\prime)}\sqrt{D_q^{(1)}(x^{\prime\prime})}}
{D_q^{(1)}(x^{\prime})+D_q^{(1)}(x^{\prime\prime})}\sqrt{q_v^+(x^\prime)}
\sqrt{q_v^+(x^{\prime\prime})}e^{-t_q^{(11)}(x^\prime,x^{\prime\prime})}\displaybreak[0]\nonumber\\ &+\frac{2D_q^{(2)}(x^\prime)D_q^{(2)}(x^{\prime\prime})}{\big[D_q^{(2)}(x^{\prime})
+D_q^{(2)}(x^{\prime\prime})\big]^2}\sqrt{q_v^-(x^\prime)}\sqrt{q_v^-(x^{\prime\prime})}
e^{-t_q^{(22)}(x^\prime,x^{\prime\prime})}\big[1-t_q^{(22)}(x^\prime,x^{\prime\prime})\big]\bigg\},
	\label{eq:F110}\displaybreak[0]\\	F_{1,2}^{q(1)}=&\frac{\xi\sqrt{1-\xi^2}}{8(1-x)}\frac{4D_q^{(2)}(x^{\prime})D_q^{(2)}(x^{\prime\prime})
+\big[D_q^{(2)}(x^{\prime})(-1+\xi)+D_q^{(2)}(x^{\prime\prime})(1+\xi)\big]^2}{D_q^{(2)}(x^{\prime})
D_q^{(2)}(x^{\prime\prime})\big[D_q^{(2)}(x^{\prime})+D_q^{(2)}(x^{\prime\prime})\big]}
\sqrt{q_v^-(x^\prime)}\sqrt{q_v^-(x^{\prime\prime})}t_q^{(22)}(x^\prime,x^{\prime\prime})
e^{-t_q^{(22)}(x^\prime,x^{\prime\prime})}\displaybreak[0]\nonumber\\
	&+\frac{c_q}{2\sqrt{2(1-\xi^2)}}\nonumber\\ &\times\bigg\{\frac{4D_q^{(1)}(x^{\prime})D_q^{(2)}(x^{\prime\prime})
+\big[D_q^{(1)}(x^{\prime})(-1+\xi)+D_q^{(2)}(x^{\prime\prime})(1+\xi)\big]^2
t_q^{(12)}(x^\prime,x^{\prime\prime})}{\sqrt{D_q^{(1)}(x^{\prime})}\big[D_q^{(1)}(x^{\prime})
+D_q^{(2)}(x^{\prime\prime})\big]^2}\sqrt{q_v^+(x^\prime)}\sqrt{q_v^-(x^{\prime\prime})}
e^{-t_q^{(12)}(x^\prime,x^{\prime\prime})}\displaybreak[0]\nonumber\\
 &-\frac{4D_q^{(2)}(x^{\prime})D_q^{(1)}(x^{\prime\prime})+\big[D_q^{(2)}(x^{\prime})(-1+\xi)
+D_q^{(1)}(x^{\prime\prime})(1+\xi)\big]^2t_q^{(21)}(x^\prime,x^{\prime\prime})}
{\sqrt{D_q^{(1)}(x^{\prime\prime})}\big[D_q^{(2)}(x^{\prime})+D_q^{(1)}(x^{\prime\prime})\big]^2}
\sqrt{q_v^-(x^\prime)}\sqrt{q_v^+(x^{\prime\prime})}e^{-t_q^{(21)}(x^\prime,x^{\prime\prime})}\bigg\},
\displaybreak[0]\\
F_{1,4}^{q(1)}=&\frac{\sqrt{q_v^-(x^\prime)}\sqrt{q_v^-(x^{\prime\prime})}(-1+x)
e^{-t_q^{(22)}(x^\prime,x^{\prime\prime})}}{\big[D_q^{(2)}(x^\prime)+D_q^{(2)}(x^{\prime\prime})\big]^2
\sqrt{1-\xi^2}}\displaybreak[0]\nonumber\\ &\times\big\{\big[D_q^{(2)}(x^\prime)(-1+\xi)+D_q^{(2)}(x^{\prime\prime})(1+\xi)\big]^2
t_q^{(22)}(x^\prime,x^{\prime\prime})+4D_q^{(2)}(x^\prime)D_q^{(2)}(x^{\prime\prime})\big\},
\end{align} for the unpolarized quark, and
\begin{align}
G_{1,1}^{q(1)}=&F_{1,4}^{q(1)},\displaybreak[0]\\ G_{1,2}^{q(1)}=&\frac{\sqrt{1-\xi^2}}{8(-1+x)}\frac{4D_q^{(2)}(x^{\prime})D_q^{(2)}(x^{\prime\prime})
+\big[D_q^{(2)}(x^{\prime})(-1+\xi)+D_q^{(2)}(x^{\prime\prime})(1+\xi)\big]^2}{D_q^{(2)}(x^{\prime})
D_q^{(2)}(x^{\prime\prime})\big[D_q^{(2)}(x^{\prime})+D_q^{(2)}(x^{\prime\prime})\big]}\sqrt{q_v^-(x^\prime)}
\sqrt{q_v^-(x^{\prime\prime})}t_q^{(22)}(x^\prime,x^{\prime\prime})e^{-t_q^{(22)}(x^\prime,x^{\prime\prime})}
\displaybreak[0]\nonumber\\
	&+\frac{c_q}{2\sqrt{2(1-\xi^2)}}\displaybreak[0]\nonumber\\ &\times\bigg\{\frac{4D_q^{(1)}(x^{\prime})D_q^{(2)}(x^{\prime\prime})+\big[D_q^{(1)}(x^{\prime})(-1+\xi)
+D_q^{(2)}(x^{\prime\prime})(1+\xi)\big]^2t_q^{(12)}(x^\prime,x^{\prime\prime})}{\sqrt{D_q^{(1)}(x^{\prime})}
\big[D_q^{(1)}(x^{\prime})+D_q^{(2)}(x^{\prime\prime})\big]^2}\sqrt{q_v^+(x^\prime)}\sqrt{q_v^-(x^{\prime\prime})}
e^{-t_q^{(12)}(x^\prime,x^{\prime\prime})}\displaybreak[0]\nonumber\\ &+\frac{4D_q^{(2)}(x^{\prime})D_q^{(1)}(x^{\prime\prime})+\big[D_q^{(2)}(x^{\prime})(-1+\xi)
+D_q^{(1)}(x^{\prime\prime})(1+\xi)\big]^2t_q^{(21)}(x^\prime,x^{\prime\prime})}{\sqrt{D_q^{(1)}(x^{\prime\prime})}
\big[D_q^{(2)}(x^{\prime})+D_q^{(1)}(x^{\prime\prime})\big]^2}\sqrt{q_v^-(x^\prime)}\sqrt{q_v^+(x^{\prime\prime})}
e^{-t_q^{(21)}(x^\prime,x^{\prime\prime})}\bigg\},\displaybreak[0]\\ G_{1,4}^{q(0)}=&2\sqrt{1-\xi^2}\bigg\{\frac{\sqrt{D_q^{(1)}(x^\prime)}\sqrt{D_q^{(1)}(x^{\prime\prime})}}
{D_q^{(1)}(x^{\prime})+D_q^{(1)}(x^{\prime\prime})}\sqrt{q_v^+(x^\prime)}\sqrt{q_v^+(x^{\prime\prime})}
e^{-t_q^{(11)}(x^\prime,x^{\prime\prime})}\displaybreak[0]\nonumber\\
	&-\frac{2D_q^{(2)}(x^\prime)D_q^{(2)}(x^{\prime\prime})}{\big[D_q^{(2)}(x^{\prime})
+D_q^{(2)}(x^{\prime\prime})\big]^2}\sqrt{q_v^-(x^\prime)}\sqrt{q_v^-(x^{\prime\prime})}
e^{-t_q^{(22)}(x^\prime,x^{\prime\prime})}\big[1+t_q^{(22)}(x^\prime,x^{\prime\prime})\big]\bigg\},
	\label{eq:G140}
\end{align}
for the longitudinally polarized quark, where
\begin{align}
	f^{q(n)}_{1,m}(x,\xi,\bm{\Delta}_\perp)&=\int d^2\bm{k}_\perp\left(\frac{\bm{k}_\perp^2}{M^2}\right)^n f^q_{1,m}(x,\xi,\bm{k}_\perp^2,\bm{\Delta}_\perp^2,\bm{k}_\perp\cdot\bm{\Delta}_\perp),\\
	t_q^{(ij)}(x^\prime,x^{\prime\prime})&=\frac{\bm{\Delta}_\perp^2}{2\kappa^2}\frac{D_q^{(i)}(x^\prime)
D_q^{(j)}(x^{\prime\prime})}{D_q^{(i)}(x^\prime)+D_q^{(j)}(x^{\prime\prime})}\frac{(1-x)^2}{(1-\xi^2)^2},\\
	F_{1,1}^{q(0)}(x,\xi,\bm{\Delta}_\perp)&=H^q_v(x,\xi,t),\\
	G_{1,4}^{q(0)}(x,\xi,\bm{\Delta}_\perp)&=\tilde{H}^q_v(x,\xi,t).
\end{align}

\subsection{Numerical results for $\langle \sin (2\phi)\rangle$}

\begin{figure}
	\centering
	\includegraphics[width=0.43\columnwidth]{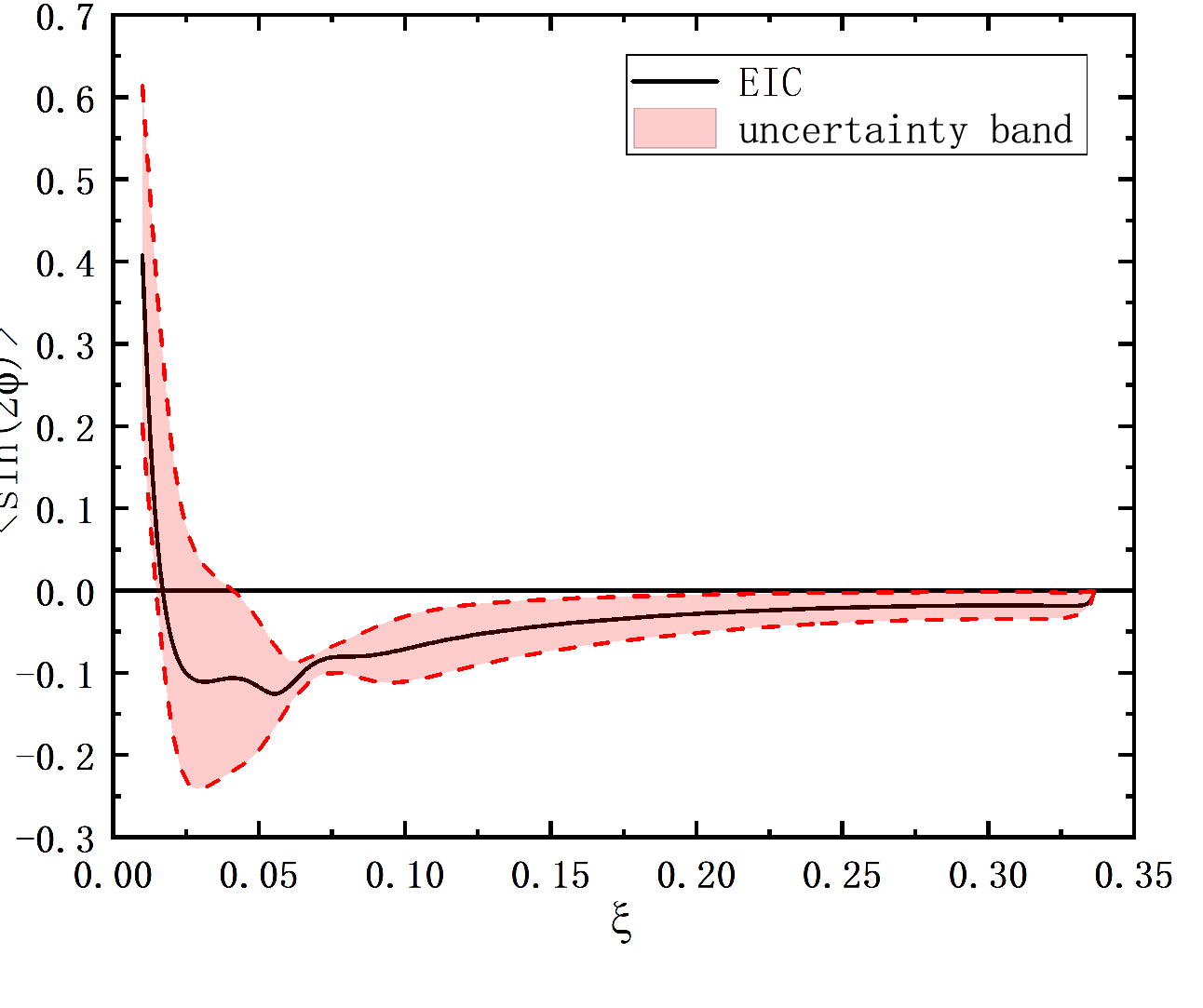}
	\includegraphics[width=0.43\columnwidth]{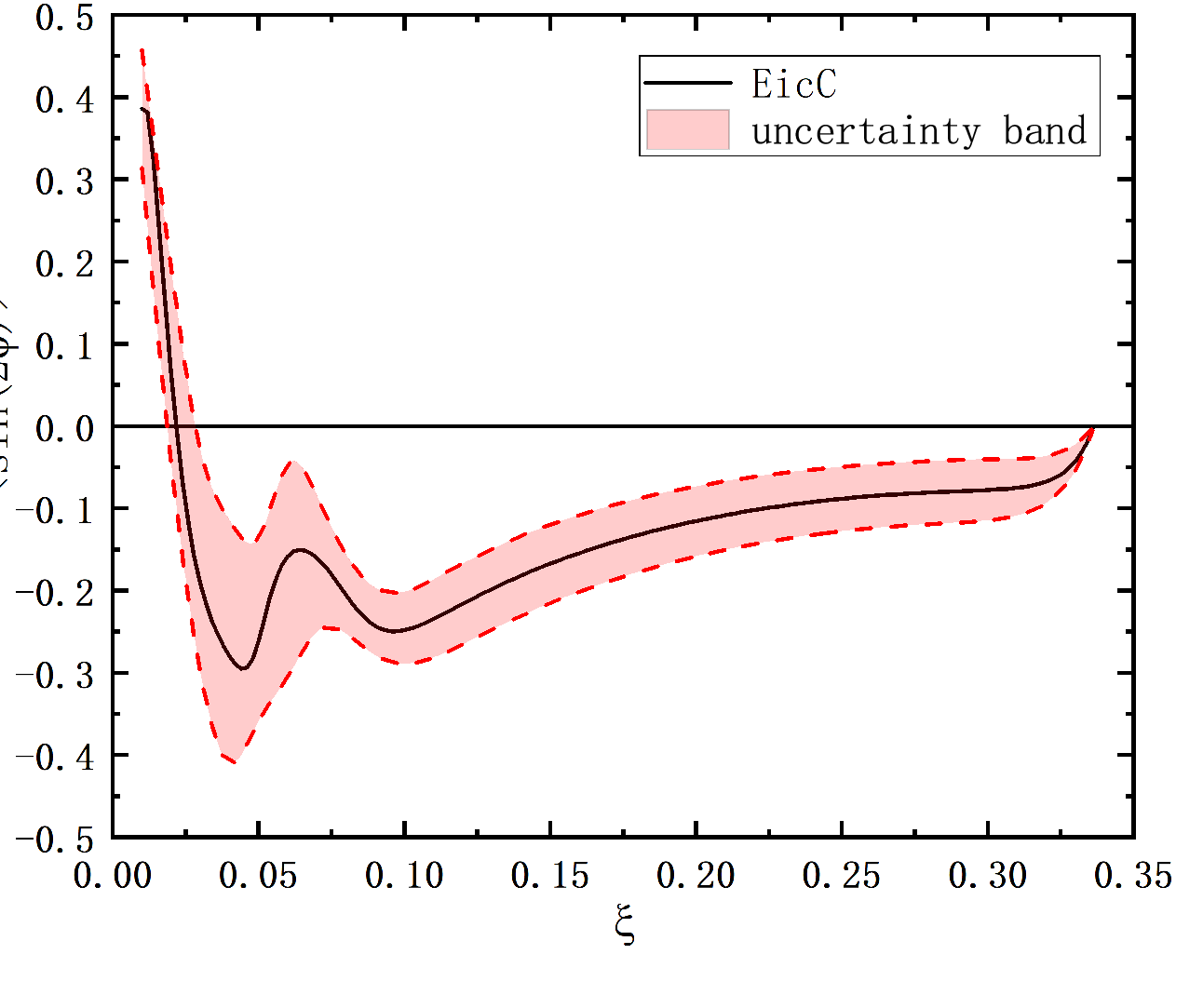}\\
	\caption{The average value of the azimuthal angular correlation $\sin (2\phi)$ as a function of $\xi$ for the EIC (left panel) and EicC (right panel) kinematics after integrating over $\phi$ and $t$.}
	\label{fig:<sin>}
\end{figure}

We now calculate the numerical results of the asymmetry for the EIC and EicC kinematics, which is defined as the average value of the $\sin(2\phi)$ azimuthal angular correlation~\cite{Bhattacharya:2023hbq}:
\begin{align}
	\langle \text{sin}(2\phi)\rangle=\frac{\int\frac{d\Delta\sigma}{d\mathcal{P}.\mathcal{S}.}\text{sin}(2\phi)
d\mathcal{P}.\mathcal{S}.}{\int\frac{d\sigma}{d\mathcal{P}.\mathcal{S}.}d\mathcal{P}.\mathcal{S}.},
\end{align}
where $d\Delta\sigma=\sigma(\lambda=1)-\sigma(\lambda=-1)$, and $d\mathcal{P}.\mathcal{S}.=d\phi dt$. The EIC and EicC kinematic correspond to $\sqrt{s_{ep}}\sim 20-140\,\text{GeV}$, $Q^2\sim 1-10^4\,\text{GeV}^2$, $x\sim 10^{-4}-1$, $y\sim 0.01-0.95$~\cite{AbdulKhalek:2021gbh,Accardi:2012qut} and $\sqrt{s_{ep}}\sim 15-20\,\text{GeV}$, $Q^2\sim 1-10^2\,\text{GeV}^2$, $x\sim 10^{-3}-0.7$, $y\sim 0.01-0.95$~\cite{Anderle:2021wcy}, respectively. We find that the real part of the quark GTMD $F_{1,4}$ in the DGLAP region associated with the canonical OAM indeed leaves a distinct signal in the above expression. 
After considering the kinematic regions corresponding to the highest luminosity of the two respective detectors and integrating $\phi$ over $[0,2\pi]$ and $t$ over $[-0.5,\,-4\xi^2M^2/(1-\xi^2)]$, we show the dependence of the asymmetry on $\xi$ in Fig.~\ref{fig:<sin>}. The left panel is the result for EIC kinematics with $Q^2=10\,\text{GeV}^2$ and $\sqrt{s_{ep}}=100\,\text{GeV}$, and the right panel is the result for EicC kinematics with $Q^2=3\,\text{GeV}^2$ and $\sqrt{s_{ep}}=16\,\text{GeV}$, where the bands with red dashed borders represent the uncertainty bands of our model results, and the black solid lines represent the results of the center values of parameters. We observe that the asymmetries are sizable and exhibit similar shapes for the EIC and EicC kinematics. 
When $\xi$ is not very small, both asymmetries are negative with the amplitude for EicC kinematics larger than that for EIC kinematics, and the most significant contributions are concentrated in the region of $0.025<\xi<0.15$. our numerical results further confirm that $\xi\sim 0.1$ is a necessary prerequisite for the $\sin(2\phi)$ azimuthal asymmetry in exclusive $\pi^0$ production as an ideal probe for detecting the quark canonical OAM in the proton.

Finally, the expected event yield for this reaction is given by
\begin{align}
N = L \sigma ,
\end{align}
where $L$ denotes the integrated luminosity and $\sigma$ the total production cross section. For electron–proton collisions at the EIC, a total integrated luminosity of $10~\text{fb}^{-1}$ is anticipated, corresponding to approximately 30 weeks of accelerator operation~\cite{AbdulKhalek:2021gbh}. By contrast, the EicC is projected to accumulate $50.5~\text{fb}^{-1}$ under an $80\%$ annual beam-on-target duty cycle~\cite{Anderle:2021wcy}. Using these luminosity values, we estimate roughly $2.68\times10^{7}$ total generated events for this process at the EIC and $6.74\times10^{7}$ at the EicC. These event counts correspond to yields integrated over the full phase space. The actual usable physics events will be significantly reduced due to limited detector acceptance (typically $0.2$–$0.5$) and reconstruction efficiencies for scattered electrons, protons, and $\pi^0$ (which decays to two photons with a reconstruction efficiency of $\sim 0.6$–$0.8$).

\section{Summary}\label{Sec:5}
In this paper, we studied the asymmetry associated with the $\sin(2\phi)$ azimuthal angular correlation in the exclusive $\pi^0$ production from electron-proton collisions, where the proton target is longitudinally polarized. This process is the clean and sensitive probe providing access to the quark GTMD $F_{1,4}$ in the DGLAP region, which establishes a direct link to the quark canonical OAM. 
We adopted a light-front quark-scalar-diquark model in which the proton LFWFs derived from the soft-wall AdS/QCD framework. 
The $x$-dependence of LFWFs is encoded in the helicity-independent and helicity-dependent valence quark PDFs, as well as the four longitudinal wave functions. 
The former was determined by simultaneously fitting the parameterized quark PDFs $q_v(x)$ and $\Delta q_v(x)$ at the initial scale $\mu_0=2\,\text{GeV}$, and the latter was determined by simultaneously fitting the quark Dirac and Pauli form factors. 
Then we calculated several angular momentum properties of valence quarks and compared them with the results from LCCQM and LC$\mathcal{X}$QSM. 
Finally, we numerically analyzed the azimuthal asymmetries $\langle\sin(2\phi)\rangle$ for the EIC and EicC kinematics and found that they are both sizable. 
Our numerical results indicated that this asymmetry indeed stands out as a promising signal for probing the quark canonical OAM in the proton and is expected to be precisely measured in future EIC and EicC.

\section*{Acknowledgements}
This work is partially supported by the National Natural Science Foundation of China under grant number 12447136 and 12150013.

\end{document}